\documentclass[%
 reprint,
 amsmath,amssymb,
 aps,
aps,prl,twocolumn,superscriptaddress
]{revtex4-2}
\usepackage{natbib}

\usepackage{graphicx}
\usepackage{dcolumn}
\usepackage{bm}
\usepackage{xcolor}
\usepackage{rotating}

\usepackage{multirow}  
\usepackage{makecell}
\usepackage{subfigure}  

\usepackage[colorlinks,
            linkcolor=red,
            anchorcolor=blue,
            citecolor=green
            ]{hyperref} 

\begin{document}

\preprint{APS/123-QED}

\title{Bayesian Analysis of Wave-Optics Gravitationally Lensed Massive Black Hole Binaries with Space-Based Gravitational Wave Detector}

\author{Yong Yuan}
\affiliation{%
 Center for Gravitational Wave Experiment, National Microgravity Laboratory, Institute of Mechanics, Chinese Academy of Sciences, Beijing, China\\
}%
\author{Minghui Du}%
 \email{duminghui@imech.ac.cn}
\affiliation{%
 Center for Gravitational Wave Experiment, National Microgravity Laboratory, Institute of Mechanics, Chinese Academy of Sciences, Beijing, China\\
}%
\author{Xin-yi Lin}%
\affiliation{%
Department of Astronomy, Beijing Normal University, Beijing 100875, China\\
}%
\author{Peng Xu}
\affiliation{%
Center for Gravitational Wave Experiment, National Microgravity Laboratory, Institute of Mechanics, Chinese Academy of Sciences, Beijing, China\\
}%

\author{Xilong Fan}
 \email{xilong.fan@whu.edu.cn}
\affiliation{
School of Physics Science And Technology, Wuhan University, No.299 Bayi Road, Wuhan, Hubei, China\\
}%

\date{\today}

\begin{abstract}

Within a Bayesian statistical framework, we jointly estimate the source and lens parameters and evaluate the relative evidence between the lensed and unlensed models. This work focuses on the wave-optics effects induced by a point-mass (PM) lens on gravitational waves (GW) from equal-mass massive binary black holes (MBHB), and assesses the capability of the space-based GW detector Taiji to detect such effects. Specifically, we investigate the impact of the redshifted lens mass $M_{Lz} \in [3\times10^5,3\times10^7] M_\odot$, impact parameter $y \in [10,50]$, source redshift $z_s \in [4,6]$, and total source mass $M_s \in [10^5,10^7] M_\odot$ on parameter estimation and model selection.
Our results show that, for the cases we studied, larger $M_{Lz}$ increases the waveform mismatch $\mathcal{MM}$, which directly enhances the waveform difference and the corresponding signal-to-noise ratio (SNR), thereby improving the ability to discriminate between the lensed and unlensed models. In contrast, for $y \geq 50$, both $\mathcal{MM}$ and SNR are too small to allow effective model discrimination in these cases. Parameter estimation further indicates that for $y < 50$, the degeneracy between $d_L$ and $M_{Lz}$ can be effectively broken. Although the Bayes factor decreases as $z_s$ increases, lensing signatures remain identifiable up to $z_s = 6$. The role of $M_s$ depends on the overlap of the GW signal with the detector’s sensitive band. Overall, effective model discrimination requires $\mathcal{MM} \gtrsim 10^{-7}$ (corresponding to SNR $>5$). 

\end{abstract}

\maketitle


\section{\label{sec:intro}1. Introduction}
Gravitational lensing occurs when electromagnetic waves pass near a massive object, causing deflection, time delay, and magnification \cite{Meylan2006glsw}. This effect has wide-ranging applications in cosmology, large-scale structure studies, exoplanet detection, dark matter research, and other areas of astrophysics. Similarly, GW can also be lensed during their propagation \cite{Ohanian1974, Deguchi1986PhRvD, Wang1996PhRvL, Nakamura1998PhRvL, Takahashi2003ApJ}. Observations and analyses of lensed GW signals provide a unique opportunity to probe the nature and distribution of dark matter, test the properties of GW, and offer new avenues for measuring cosmological parameters and exploring large-scale structures in the universe \cite{Fan2017PhRvL, Liao2018ApJ, Tao2019ApJ, Hannuksela2020MNRAS, Sereno2011MNRAS, Liao2017NatCo, Cao2019NatSR, Li2019ApJ, Hai2020MNRAS, Zhou2023MNRAS, Urrutia2022MNRAS, Chung2021PhRvD, Gais2022ApJ, Broadhurst2020arXiv, Gao2022MNRAS, Liu2019MNRAS, Lin2023PhRvD, Lin2025arXiv}.

LIGO-Virgo-KAGRA Collaboration have already detected a large number of GW events from compact binary coalescences, including binary black hole mergers, black hole–neutron star mergers, and binary neutron star mergers. In 2017, the first GW event with an electromagnetic counterpart was observed \cite{Abbott2017PhRvL,Covino2017NatAs, GoldsteinApJL2017, Kasen2017Natur, Savchenko2017ApJL, Zhang2018NatCo}.  By the end of the O4a observing run, approximately 200 GW events had been detected \cite{Abbott2023PhRvX, LIGO2025arXiv1, LIGO2025arXiv2, LIGO2025arXiv3}. Although no lensed GW signals have been observed so far, extensive studies have been carried out on the theoretical and methodological aspects of lensed GW \cite{Fan2017PhRvL, Broadhurst2019arXiv, McIsaac2020PhRvD, Hannuksela2019ApJ, Liu2021ApJ, Abbott2021ApJ, Baker2017PhRvD, Srashti2021PhRvD, Lai2018PhRvD, Diego2020PhRvD, Xu2022ApJ, Abbott2024ApJ}. Moreover, it is predicted that third-generation ground-based detectors, such as the Cosmic Explorer \cite{Abbott2017CQGra} and the Einstein Telescope \cite{Punturo2010CQGra}, will be able to detect lensed GW events.

Space-based GW detectors, such as LISA \cite{Pau2017arXiv}, Taiji \cite{Hu2017NSRev}, and TianQin \cite{Luo2016CQGra}, are planned future missions designed to detect GW in the milli-Hertz (mHz) frequency band. One of their primary scientific objectives is to observe the mergers of MBHB, typically in the mass range $10^4 \sim 10^7 M_\odot$ \cite{Pau2017arXiv}. Owing to their exceptional sensitivity, these detectors are expected to detect hundreds to thousands of MBHB merger events up to redshifts as high as $z \sim 20$, with SNR reaching $10^2 - 10^3$ over their mission lifetimes \cite{Klein2016PhRvD, Diao2025arXiv, Ruan2021Resea, Wang2019PhRvD}. At such high redshifts, gravitational lensing by large-scale structures can no longer be neglected \cite{Takahashi2006ApJ, Yoo2007ApJ}. Furthermore, since GWs emitted by MBHB occur at much lower frequencies compared to those from stellar-mass binaries, this frequency regime opens the possibility of probing wave-optics effects, such as diffraction, in the lensed signals. In particular, when the Schwarzschild radius of the lensing object is smaller than the GW wavelength $\lambda$, diffraction effects become decisive in shaping the observed waveform. For diffraction to play a significant role, the lens mass $M_L$ must satisfy the condition \cite{Takahashi2003ApJ, Mesut2023PhRvD}
\begin{equation}
M_{{L}}\lesssim10^{5}M_{\odot}\left(\frac{f}{\mathrm{Hz}}\right)^{-1},
\end{equation}

Wave-optics effects can induce frequency-dependent amplitude and phase modulations in GW signals \cite{Takahashi2003ApJ}.
Consequently, the detection of such lensing-induced features by space-based GW detectors can be used to infer lens parameters, such as the redshifted lens mass $M_{Lz} = (1+z_L)M_L$ (with $z_L$ denoting the lens redshift), as well as the source position in the source plane. Furthermore, if the event rate is sufficiently high, the measurement of lens parameters may not only reveal the overall distribution of lensing objects and refine existing lens models, but also provide potential new avenues for probing the nature of dark matter \cite{Mesut2023PhRvD}.

The wave-optics effects in gravitational lensing of GWs have been extensively studied in the literature \cite{Ohanian1974,Nakamura1998PhRvL,Takahashi2003ApJ,Oguri2020ApJ,Cremonese2021PhRvD, Gao2022MNRAS, Paolis2002AA,Takahashi2004AA,Takahashi2006ApJ, Meeha2020MNRAS, Gao2022MNRAS, Mesut2023PhRvD}. In the study by Takahashi and Nakamura \cite{Takahashi2003ApJ}, GW lensed by either PM or singular isothermal sphere (SIS) lenses were considered, with lens masses in the range $M_{Lz} \in [10^6, 10^9] M_\odot$. The results showed that, for space-based gravitational wave detectors operating in the mHz band, when the total mass of MBHB is $M_{s} = 2 \times 10^6 M_\odot$ with a mass ratio $q = 1$, wave-optics effects enable accurate measurement of SIS lens parameters in the range $M_{Lz} \approx 10^6 - 10^8 M_\odot$. Notably, subsequent studies \cite{Gao2022MNRAS, Mesut2023PhRvD} demonstrated that wave-optics effects of lensed GW could still be detectable even at larger impact parameters and for lower lens masses. This result extends the detectable parameter space beyond that considered in earlier work.

For space-based GW detectors, the lenses responsible for diffraction effects are primarily low-mass dark matter (DM) haloes and subhaloes within massive main haloes (\cite{Takahashi2003ApJ, Takahashi2004AA}. However, the cold dark matter (CDM) model predicts that the most abundant haloes should lie in the mass range $10^6 \sim 10^9 M_\odot$ (e.g., \cite{Cooray2002PhR, Han2016MNRAS}). Notably, \cite{Takahashi2003ApJ} imposed a maximum impact parameter of $y = 3$ to ensure strong and detectable diffraction effects. This condition, however, may be overly conservative. Owing to the high signal-to-noise ratios (SNR) of space-based GW sources, weaker lensing signals could still be detectable at larger impact parameters. Therefore, by including these low-mass haloes and relaxing the restriction on $y$, the detection rate of lensed gravitational wave signals could be significantly higher than previously estimated.

Extensive studies have been conducted on the detection of GW lensing effects, the estimation of lensing parameters, and the discrimination between different lens models \cite{Cao2014PhRvD, Sun2019arXiv,  Cusin2021MNRAS, Gais2022ApJL, Wright2022ApJ, Chen2024PhRvD, Cremonese2023AnP, Mishra2024MNRAS, Nerin2025PhRvD, Gao2022MNRAS, Mesut2023PhRvD, Lin2023PhRvD, Lin2025arXiv}.
\cite{Cao2014PhRvD} employed the Fisher information matrix to investigate the impact of PM and SIS lens models on parameter estimation with ground-based GW detectors. \cite{Wright2022ApJ, Gais2022ApJL, Chen2024PhRvD} adopted a Bayesian statistical framework to jointly infer both source and lens parameters and further assessed the distinguishability between the PM and SIS models. Their results indicate that breaking the degeneracy between luminosity distance and lens parameters is only feasible when the impact parameter is small ($y<3$), i.e., when the lensing magnification factor is sufficiently large. This limitation arises because ground-based detectors observe compact binary coalescences over relatively short durations, allowing lensing and unlensed signals to be distinguished only when the lensing effect is pronounced. In contrast, space-based GW detectors enable long-term tracking of GW sources, thereby accumulating differences between lensed and unlensed waveforms, which could alleviate this degeneracy.
In addition, \cite{Mishra2024MNRAS} investigated the parameter estimation bias induced by lensing effects and further explored the population properties and detection rates in the presence of multiple lensed GW events. \cite{Gao2022MNRAS, Mesut2023PhRvD} analyzed the prospects for detecting lensed GW events with space-based detectors. \cite{Lin2023PhRvD, Lin2025arXiv} combined Fisher information matrix analyses with Bayesian inference to study the impact of lensing on parameter estimation and discussed model selection under different lensing scenarios.

Building upon previous studies, this work focuses on exploring DM haloes and subhaloes, or CDM, as potential lenses for GW generated by MBHBs. We adopt the PM lens model to characterize the lensing effect, concentrating on the following parameter space: the total binary mass is set to $10^5 \sim 10^7 M_\odot$, with source redshifts in the range $z \in [4,6]$; the redshifted lens mass is chosen as $M_{Lz} \in [3\times10^5, 3\times10^7] M_\odot$, and the dimensionless impact parameter lies within $y \in [10, 150]$. Under these conditions, we simulate lensed GW signals from MBHB mergers based on the sensitivity of the Taiji detector. The central question addressed in this study is whether the lensing signatures can be sufficiently distinct to statistically discriminate between lensed and unlensed models within the above parameter space. This investigation carries significant physical implications: if lensing effects can indeed be identified, they will not only provide an independent astrophysical probe of the mass distribution of DM haloes and subhaloes but also serve as a novel observational test for CDM models. Moreover, such findings would highlight the broader potential of space-based GW detectors in advancing both cosmology and dark matter research.

The paper is organized as follow. In Section 2.1, we provide a brief overview of the PM lensing model. Section 2.2 introduces the Bayesian statistical framework employed in this study. In Section 3.1, we describe the construction and simulation of lensed GW signals, followed by Section 3.2, where we present the results of parameter estimation and model selection. Finally, in Section 4, we summarize our main findings and discuss possible extensions and future research directions. Throughout this paper, we assume a flat $\Lambda$CDM cosmology with the parameters $\Omega_m=0.3111$, $\Omega_\Lambda=1-\Omega_m$, $H_0=67.66 \ {\rm km/s}$ \cite{Planck2020AA}.

\section{2. Method}

\subsection{2.1 Point mass}
\label{sec:PM}

\begin{figure}
\centering 
\includegraphics[width=8.5cm]{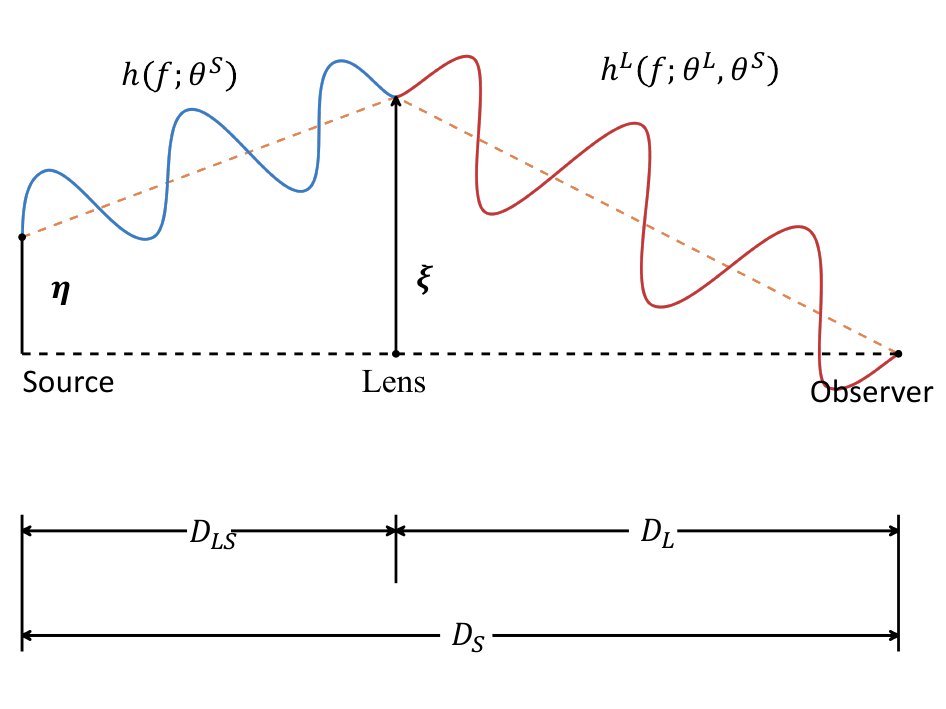}
\caption{Schematic illustration of GW lensing. The vectors $\bm\eta$ and $\bm\xi$ represent the position on the source plane and the lens plane. The angular diameter distances $D_L$, $D_S$, and $D_{LS}$ are measured in the observer's rest frame.}
\label{fig:lens_show}
\end{figure}

For simplicity, the author assume a PM profile for DM haloes and subhaloes. In this case, we can follow \cite{Takahashi2003ApJ} to model the diffraction of GWs. Using a more realistic Navarro-Frenk-White profile usually leads to a slightly smaller magnification factor. The basic picture of lensing of GWs is illustrated in FIG.~\ref{fig:lens_show}, where $D_L$, $D_S$, and $D_{LS}$ denote the angular diameter distances to the lens, to the source and their difference, respectively. All these quantities are measured in the frame of the observer.
 
The effect of a gravitational lens on the propagation of GW can be described by solving the complex-valued diffraction integral for a given source frequency $f$ \cite{Takahashi2003ApJ}
\begin{equation}
F(f,\bm{y})=\frac{D_\mathrm{S}(1+z_\mathrm{L})\xi_0^2}{cD_\mathrm{L}D_\mathrm{LS}}\frac{f}{i}\int\mathrm{d}^2\bm{x}\exp[2\pi ift_\mathrm{d}(\bm{x},\bm{y})].
\label{eq:factor}
\end{equation}
where the integral is taken over all possible paths, including those that are not geodesics. In this expression, $D_L$, $D_S$, and $D_{LS}$ denoted the angular-diameter distances from the observer to the lens, from the observer to the source, and from the lens to the source, respectively. The dimensionless vectors $\bm x$ and $\bm y$ are defined as 
\begin{equation}
\bm x = \frac{\bm\xi}{\xi_0}; \bm y = \frac{D_L}{\xi_0D_S}\bm\eta,
\end{equation}
where $\bm\xi$ and $\bm \eta$ are the physical coordinates of the image in the lens plane and of the source in the source plane, respectively.

The normalization factor $\xi_0$ is an arbitrary length scale, typically chosen to match the characteristic scale of the lensing configuration. The time delay associated with a path passing through position $\bm x$ is given by
\begin{equation}
t_\mathrm{d}(\boldsymbol{x},\boldsymbol{y})=\frac{D_\mathrm{S}\xi_0^2}{cD_\mathrm{L}D_\mathrm{LS}}(1+z_\mathrm{L})\left[\frac{1}{2}|\boldsymbol{x}-\boldsymbol{y}|^2-\psi(\boldsymbol{x})+\boldsymbol{\phi}(\boldsymbol{y})\right],
\end{equation}
where ${\psi}(\boldsymbol{x})$ is the deflection potential, and $\phi(\boldsymbol{y})$ is a source-position-dependent constant that sets the zero-point of the time delay. Since $\phi(\boldsymbol{y})$ does not affect relatively time delay between different paths. For convenience, we fix it such that the minimum of $t_\text{d}(\bm x,\bm y)$ is zero for a given $\bm y$. 

To simplify the analysis, we now restrict our attention to spherically symmetric lenses. In this case, the problem becomes one-dimensional: the deflection potential depends only on the magnitude of $\bm x$, i.e., $\psi(\bm x) = \psi(x)$, and likewise $\phi(\bm y) = \phi(y)$, where $x\equiv|x|$ and $y\equiv|y|$. Without loss of generality, we align the polar axis with $\bm y$, allowing us to perform the angular part of the integral in Eq. \eqref{eq:factor} analytically. The amplification factor then becomes 
\begin{equation}\begin{aligned}
F(w,y)= & \frac{w}{i}\mathrm{exp}\left\{iw\left[\frac{y^2}{2}+\phi(y)\right]\right\} \\
 & \times\int_0^\infty x\mathrm{d}x\exp\left\{iw{\left[\frac{x^2}{2}-\psi(x)\right]}\right\}J_0(wxy).
\end{aligned}
\label{eq:Fw}
\end{equation}
where
\begin{equation}
w\equiv\frac{D_{\mathrm{S}}\xi_{0}^{2}(1+z_{\mathrm{L}})(2\pi f)}{cD_{\mathrm{L}}D_{\mathrm{LS}}}
\label{eq:w}
\end{equation}
is the dimensional frequency, and $J_0$ denotes the zeroth-order Bessel function. 

To proceed, we adopt the PM lens model as a representative example gravitational lensing. In this model, the entire lensing mass is concentrated at a single point, with the mass density given by $\rho_\text{PM}(\bm r)=M_L\delta^3(\bm r)$, where $\delta^3(\bm r)$ is the three-dimensional Dirac delta function. A natural choice for $\xi_0$ is the Einstein radius, defined as 
\begin{equation}
\xi_0=\left(\frac{4M_LGD_LD_{LS}}{c^2 D_S}\right)^{1/2},
\end{equation}
which characterizes the typical angular scale of the lensing effect.

Under this choice of $\xi_0$, the deflection potential becomes $\psi(x) = \ln(x)$, allowing the radial integral in Eq. \eqref{eq:Fw} to be evaluated analytically. The amplification factor for the PM lens is then given by \cite{Takahashi2003ApJ}
\begin{equation}\begin{aligned}
F(w,y) & =\exp\left\{\frac{\pi w}{4}+i\frac{w}{2}\left[\ln\frac{w}{2}-2\phi(y)\right]\right\} \\
 & \times\Gamma\left(1-\frac{w}{2}i\right)_1F_1\left(\frac{w}{2}i,1;\frac{wy^2}{2}i\right).
 \label{eq:Fw}
\end{aligned}\end{equation}
where, $w=8\pi G M_L(1+z_L)f/c^3$ is the dimensionless frequency, $\phi(y)=(x_+-y)^2/2-\ln x_+$, $x_+=[(y^2+4)^{1/2}+y]/2$, $\Gamma$ is the Gamma function, and $_1F_1(a,b;z)$ is the confluent hypergeometric function which is also called complex Kummer function. The corresponding amplification factor $|F|$ and phase-change factor $\theta_F$ are
\begin{equation}
|F| = \sqrt{FF^*}, \theta_F=-i\ln[F/|F|],
\end{equation}
where $F^*$ is the complex conjugate of $F$.

\begin{figure}
\centering
\includegraphics[width=8cm]{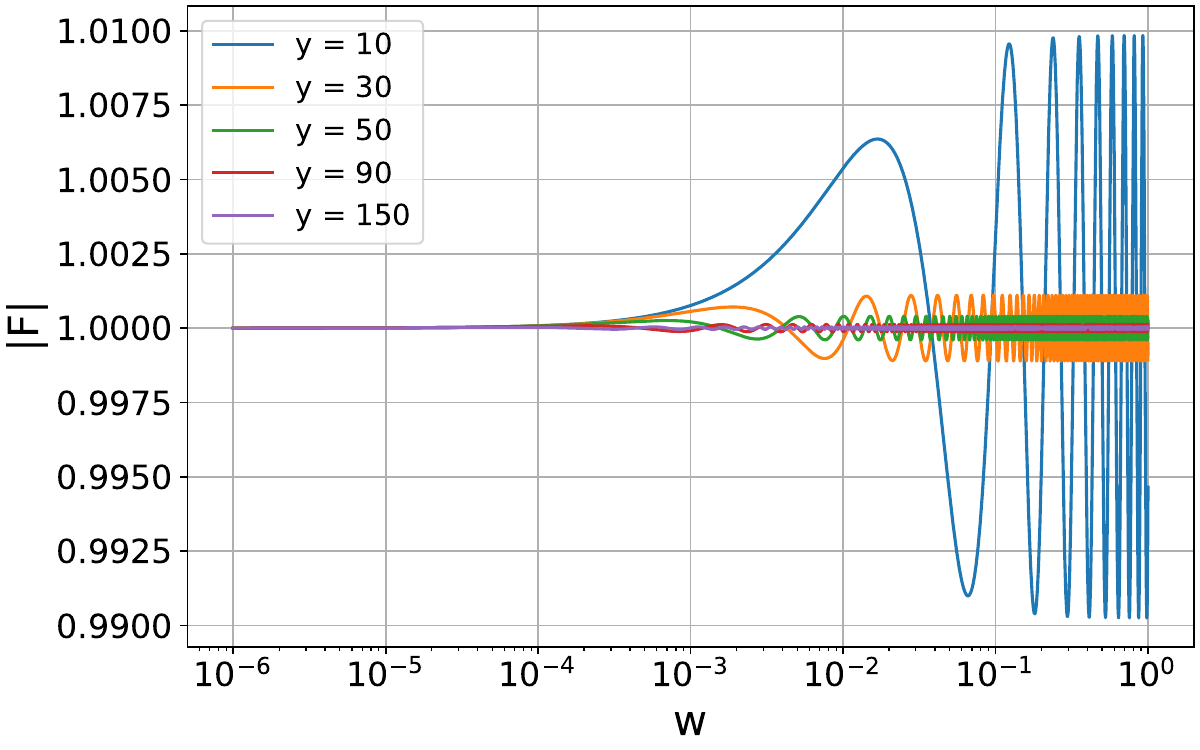}
\includegraphics[width=8cm]{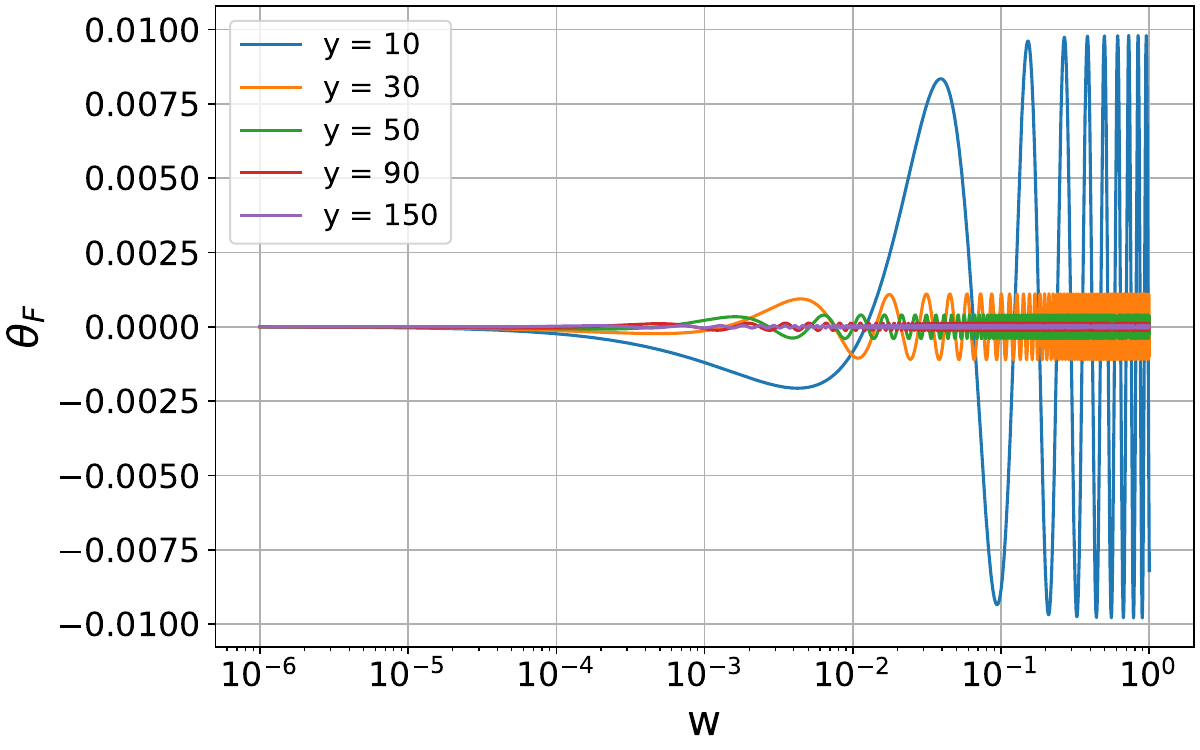}
\caption{The amplification factor (top panel) and the phase shift (bottom panel) as functions of the dimensionless frequency in the PM model. Different curves correspond to different impact parameters $y$.} 
\label{fig:PM_F}
\end{figure}

With the help of the GLoW module \cite{Hector2025PhRvD}, we efficiently compute the amplification factor $|F|$ and its phase $\theta_F$. As shown in FIG.~\ref{fig:PM_F}, both $|F|$ and $\theta_F$ depend on the dimensionless frequency $w$ and the impact parameter $y$. The amplification factor $|F|$ generally decreases with increasing $y$ and exhibits oscillations around unity for different values of $y$. Meanwhile, the phase shift $\theta_F$ remains non-zero regardless of the impact parameter. These features suggest that, even when wave-optics effects are weak, they may still be detectable if the SNR of the event is sufficiently high. Furthermore, when $y$ is fixed, both $|F|$ and $\theta_F$ can vary significantly with $w$. In particular, the critical value of $y$, above which diffraction effects become undetectable, depends on $w$. According to Eq.~(\ref{eq:w}), the dimensionless frequency $w$ depends on the lens mass, lens redshift, and the gravitational wave frequency. We will further investigate how this dependence affects parameter estimation in lensed gravitational wave signals in the following sections. Finally, as the impact parameter $y$ increases, the location of the first peak in both $|F|$ and $\theta_F$ shifts to lower values of $w$.

\subsection{2.2 Bayesian statistical framework}

We simulated GW signals from the merger of MBHB based on the Taiji configuration. The simulated GW signal without lensing can be derived from Eq. (\ref{eq:un-d})
\begin{equation}
d(f;\theta^S) = h(f;\theta^S) + n(f),
\label{eq:un-d}
\end{equation}
Here, $h(f; \theta^S)$ denotes the unlensed GW signal, and $n$ represents the detector noise. In this work, we neglect the effects of black hole spins. As a result, the source parameter vector is simplified to $\theta^S=(\mathcal{M}_c,q, d_L, t_c, \phi_c, \iota, \lambda, \beta, \psi)$, where $\mathcal{M}_c$ is the chirp mass, $q$ is the mass ratio, $d_L$ is the luminosity distance, $t_c$ is the coalescence time, $\phi_c$ is the coalescence phase, $\iota$ is the inclination angle, $\lambda$ and $\beta$ are the Ecliptic longitude and latitude of the source in the sky, and $\psi$ is the polarization angle. Since our study focuses on equal-mass massive binary black hole systems, we fix the mass ratio and exclude it from the set of free parameters. In addition, both the inclination angle and the polarization angle are treated as subdominant in this analysis and are thus also omitted from the parameter space. Accordingly, we define the set of parameters of interest as $\theta^A={(\mathcal{M}_c, d_L, t_c, \phi_c, \lambda, \beta)}$.

The simulated lensed GW signal under the PM model is obtained according to Eq. (\ref{eq:L-h})
\begin{equation}
\begin{split}
d^L(f; \theta^L, \theta^S) &= F(f;\theta^L)h(f;\theta^S) + n(f)\\
&= h^L(f;\theta^L, \theta^S) + n(f),
\end{split}
\label{eq:L-h}
\end{equation}
where $F(f;\theta^L)$ is amplification factor and $\theta^L = {(M_{Lz}, y)}$.

Within the Bayesian statistical framework, we perform a systematic analysis of parameter estimation for GW events under both lensed and unlensed scenarios, using simulated lensed GW data. The corresponding Bayesian formulations for the lensed and unlensed cases are given by Eq. (\ref{eq:B-lens}) and Eq. (\ref{eq:B-N}), respectively.
\begin{equation}
\begin{split}
p(\theta^A,\theta^L|d^L,M^L) 
= &\;\int \mathcal{L}(d^L|\theta^S,\theta^L,M^L)
       p(\theta^A,\theta^L,M^L)\, \\
&\text{d}q\,\text{d}\iota\,\text{d}\psi  \;\times \big[\mathcal{Z}^L(d^L|M^L)\big]^{-1}
\end{split}
\label{eq:B-lens}
\end{equation}

\begin{equation}
p(\theta^A|d^L,M^N)= \frac{\int \mathcal{L}(d^L|\theta^S,M^N)p(\theta^A|M^N)\text{d}q\text{d}\iota\text{d}\psi}{\mathcal{Z}^N(d^L|M^N)}
\label{eq:B-N}
\end{equation}
Here, $M^L$ and $M^N$ denote the lensed and unlensed models, respectively. $\mathcal{L}$ represents the likelihood function, $p$ denotes the prior distribution, and $\mathcal{Z}$ is the Bayesian evidence for a given model. The evidences for the lensed and unlensed models are defined as
\begin{equation}
\mathcal{Z}^L = \int \mathcal{L}(d^L|\theta^S, \theta^L,M^L)p(\theta^A,\theta^L,M^L)\text{d}\theta^S\text{d}\theta^L
\end{equation}
\begin{equation}
\mathcal{Z}^N = \int \mathcal{L}(d^L|\theta^S,M^N)p(\theta^A|M^N)\text{d}\theta^S
\end{equation}

We perform parameter estimation on the lensed data using both the lensed model and the unlensed model. By comparing the Bayesian evidences of the two models, we compute the Bayes factor
\begin{equation}
\ln\mathcal{B}^{L/N} = \ln\mathcal{Z}^L -\ln\mathcal{Z}^N,
\end{equation}
which quantifies the relative support for the lensing hypothesis.

\section{3. Data analysis}

\subsection{3.1 Data simulation and waveform analysis}

Under the PM model, we simulate lensed GW signals from MBHB according to Eq.~(\ref{eq:L-h}). The GW waveforms from MBHB are generated using the IMRPhenomD \cite{Husa2016PhRvD, Khan2016PhRvD} waveform model, which describes the dominant $\ell = 2, |m| = 2$ modes of aligned-spin binary systems. In this study, we restrict ourselves to the dominant  mode without including higher-order modes, as this simplification is sufficient for investigating the lensed GW waveforms. 
The detailed procedure for  
 simulating the time-delay interferometry (TDI) data 
is presented in the Appendix. 
Throughout this work, we exclusively use the noise-orthogonal TDI-$A_2$, $E_2$ channels for Bayesian analysis~\cite{Prince2002PhRvD}.  
While for visualization in the Figures, only the $A_2$-channel waveform is used as a representative example, hence it is convenient to define: 
\begin{equation}
h(f;\theta^S) \equiv h_{\tilde{A}_2}(f;\theta^S).
\end{equation}
During the data simulation, without loss of generality, we set the starting point to day 60 of the  mission orbit and simulate a total duration of 90 days.
To ensure that the simulated data cover the inspiral, merger, and ringdown phases of the MBHB system, the merger time is placed on the day before the end of the segment, i.e., day 149. In addition, when analyzing the case of binary systems with component masses of $10^5,M_\odot$, the sampling frequency of the space-based GW detector is set to 0.3, while for systems with component masses greater than $10^5,M_\odot$, the sampling frequency is set to 0.03. This configuration ensures both waveform accuracy and improved computational efficiency.

To begin with, we compare the differences between lensed and unlensed waveforms. The characteristic strain is selected as the metric for this comparison, since the area enclosed between the characteristic strain curve and the detector's noise curve corresponds to the commonly used SNR. This provides a more intuitive means of evaluating how lensing affects the waveform's detectability. The characteristic strain of the signal ($h_c$) and the noise ($h_n$) are defined in Eq. (\ref{eq:hc}) and Eq. (\ref{eq:hn}) \cite{Moore2015CQGra},
\begin{equation}
h_c(f)^2 = 4f^2|h(f;\theta^S)|^2,
\label{eq:hc}
\end{equation}
\begin{equation}
h_n(f)^2 = fS_n(f),
\label{eq:hn}
\end{equation}
where $S_n(f)$ is one-sided noise power spectrum density (PSD). 

\begin{figure}[h]
\centering
\includegraphics[width=8cm]{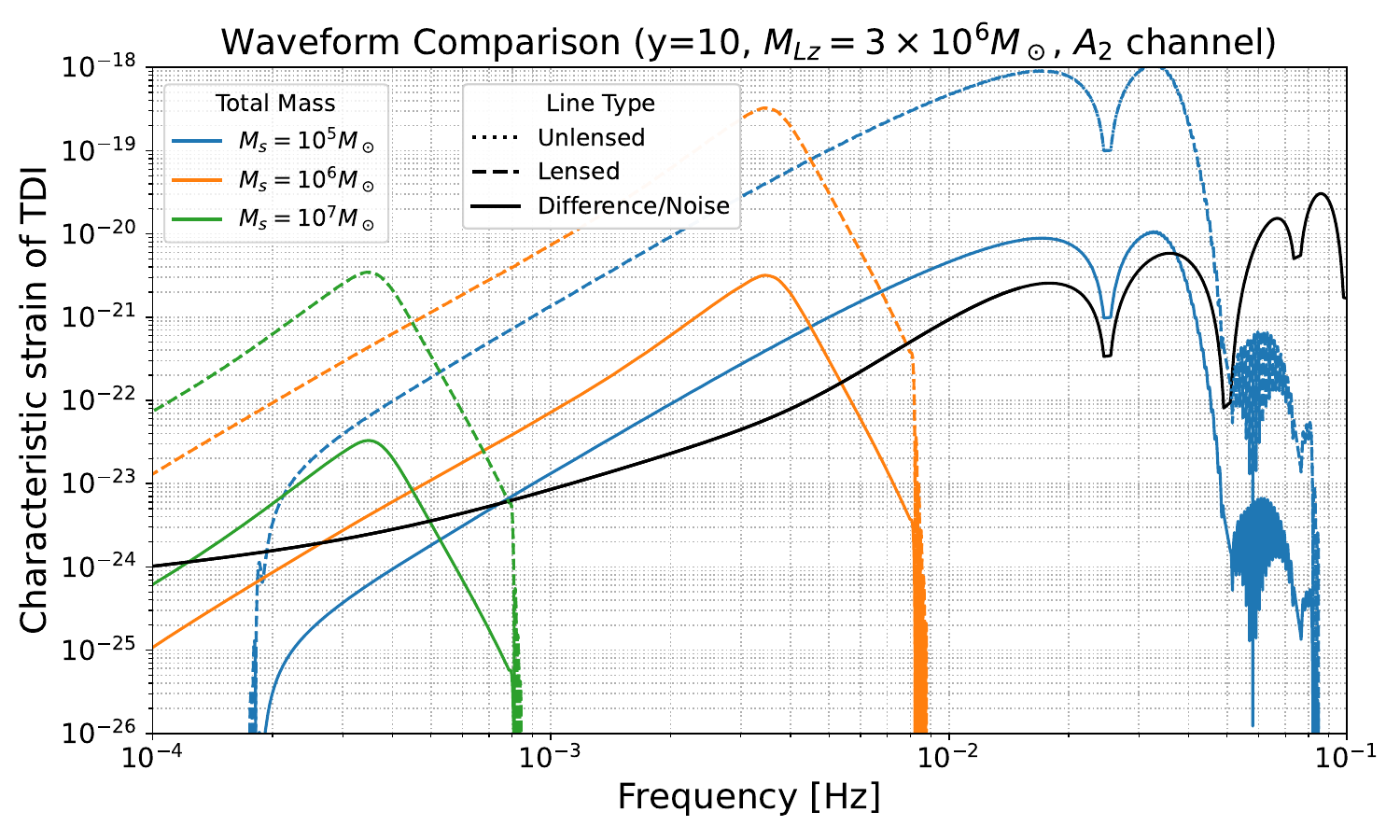}

\includegraphics[width=8cm]{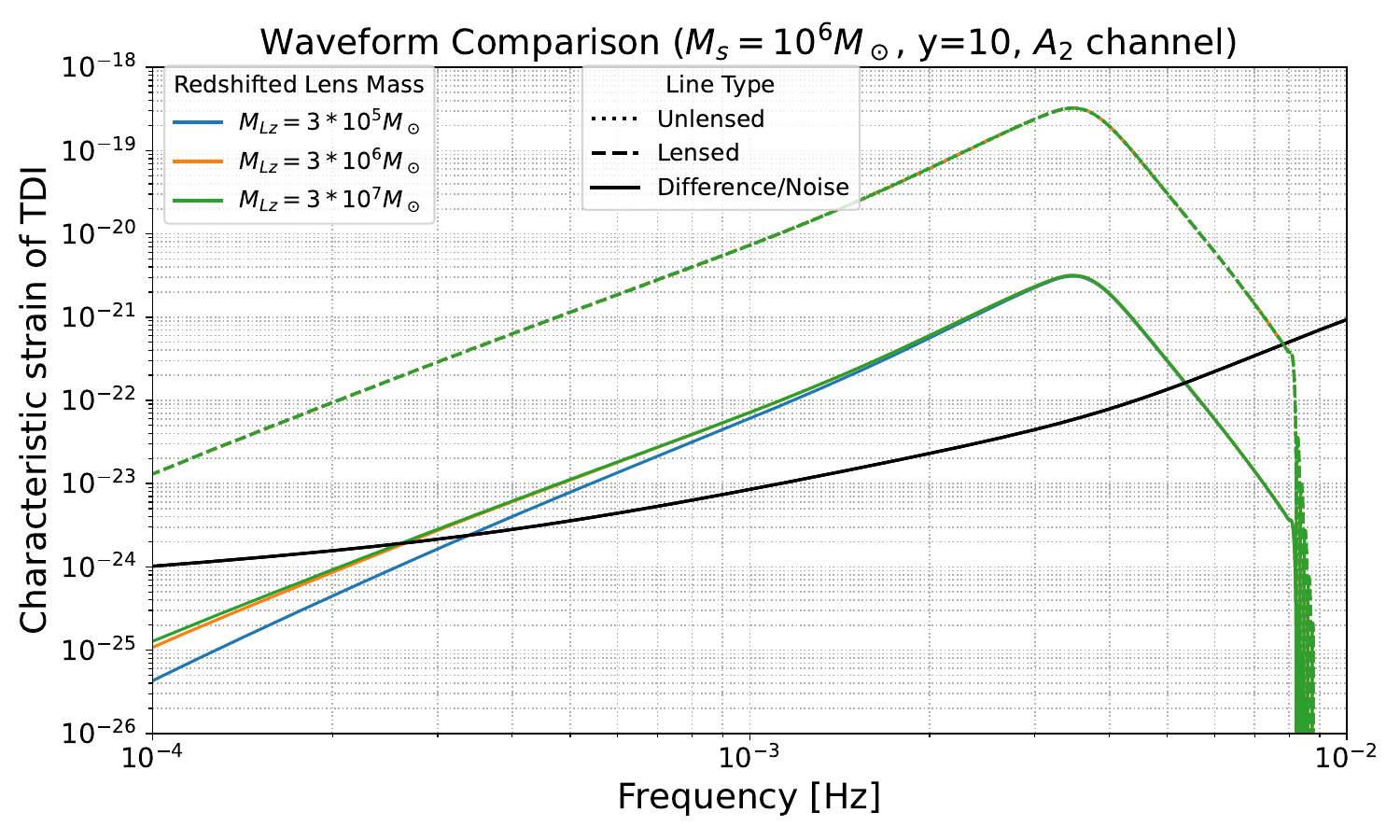}

\includegraphics[width=8cm]{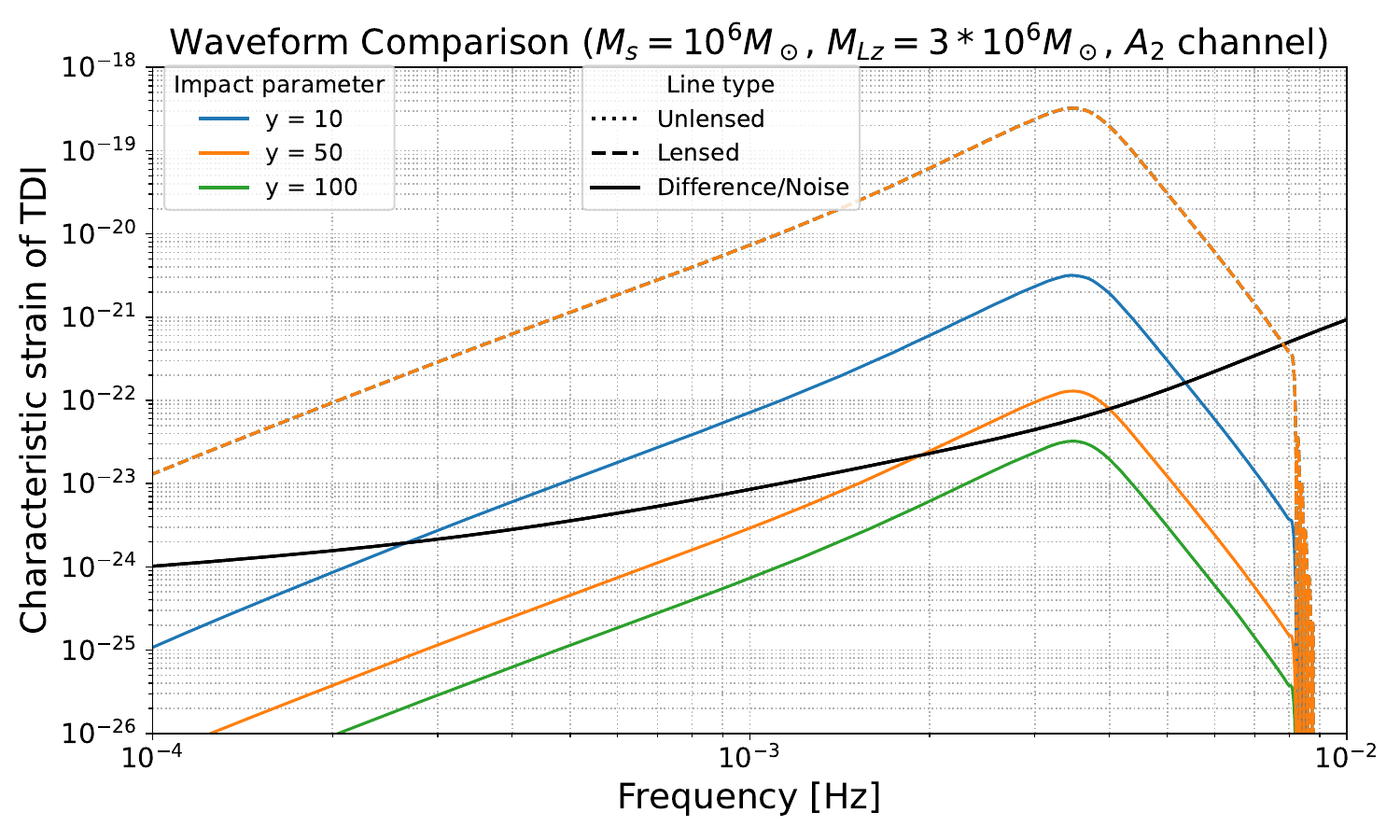}
\caption{Comparison of characteristic strain between lensed and unlensed GW signals in the frequency domain for the Taiji A channel. Top panel: $M_{Lz}=3\times10^6 M_\odot$ and $y=10$, with varying the total source-frame binary mass as $M_s = 10^5, 10^6, 10^7 M_\odot$. Middle panel: \(M_s = 10^6\,M_\odot\) and \(y = 10\), with varying redshifted lens masses \(M_{Lz}\) = \(3 \times 10^5\,M_\odot\), \(3 \times 10^6\,M_\odot\), \(3 \times 10^7\,M_\odot\). Bottom panel: $M_s$=\(10^6\,M_\odot\) and $M_{Lz}=$\(3 \times 10^6\,M_\odot\), with varying impact parameters $y$ = 10, 50, 100. Black solid line denotes the characteristic noise sensitivity curve of the detector, dashed lines represent the lensed waveforms,  colored dotted lines represent unlensed waveforms, and colored solid lines show the residuals (lensed minus unlensed).
}
\label{fig:channel}
\end{figure}

The magnification factor derived in Eq.~(\ref{eq:Fw}) is frequency-dependent. To apply it, we first compute the unlensed GW signal in the frequency domain. For clarity, illustrative examples are shown in FIG.~\ref{fig:channel}. For simplicity, all simulations assume equal-mass binaries with zero eccentricity, zero spin, zero inclination, and zero polarization, with the source redshift fixed at $z_{s} = 4$ in all cases. 
The results show that the lensed and unlensed waveforms are nearly indistinguishable overall (dashed and dotted lines nearly overlapping). However, when comparing the characteristic strain residuals of the lensed and unlensed cases (colored solid lines) with the detector’s characteristic noise sensitivity curve (black solid line), we find that the residuals remain significantly above the sensitivity curve across a range of frequencies. This demonstrates that, although the two waveforms appear almost identical to the eye, the subtle lensing-induced differences are sufficient to yield a measurable impact on the SNR.
In the top panel of FIG.~\ref{fig:channel}, we fix $M_{Lz}=3\times10^6 M_\odot$ and $y=10$, while varying the total source-frame binary mass as $M_s = 10^5, 10^6,$ and $10^7 M_{\odot}$. As expected, systems with larger total masses merge at lower frequencies, consistent with the inverse relationship between the merger frequency and the total binary mass. In addition, we observe that the GW signals converge at high frequencies, while the lensing effect becomes more pronounced at low frequencies, consistent with the trend seen in FIG.~\ref{fig:PM_F}. For the $M_s = 10^5,M_{\odot}$ system, a cutoff appears at around $9\times10^{-4}$ Hz, because the frequency corresponding to 59 days before the merger lies outside the 60-day data segment used in the simulation, resulting in the truncation. Comparing the characteristic strain residuals between lensed and unlensed waveforms (colored solid lines) with the detector’s characteristic noise sensitivity curve (black solid line) shows that, over a certain frequency range, the residuals lie significantly above the sensitivity curve, indicating that lensing-induced waveform differences are sufficient to cause a noticeable change in the SNR.
In the middle panel of FIG.~\ref{fig:channel}, we fix the total mass of the binary system to \(10^6\,M_\odot\), set \(y = 10\), while varying the redshifted lens masses \(M_{Lz}\) as \(3 \times 10^5\,M_\odot\), \(3 \times 10^6\,M_\odot\), and \(3 \times 10^7\,M_\odot\), respectively. As shown in the figure, a smaller \(M_{Lz}\) results in a smaller difference between the lensed and unlensed waveforms. This is because, at a given frequency, a smaller lens mass corresponds to a smaller dimensionless frequency \(w\). According to FIG.~\ref{fig:PM_F}, the amplification factor decreases with decreasing \(w\) (for $w<2\times10^{-2}$), reducing the strength of lensing-induced waveform modulation and thus diminishing the discrepancy between the lensed and unlensed waveforms. At higher frequencies, the waveform differences for different lens masses converge, exhibiting no significant distinction.
In the bottom panel of FIG.~\ref{fig:channel}, we fix the total mass of the binary system to \(10^6\,M_\odot\) and the redshifted lens mass to \(3 \times 10^6\,M_\odot\), while varying the source impact parameter \(y\) as 10, 50, and 100. As shown in the figure, the difference between the lensed and unlensed waveforms decreases with increasing \(y\), consistent with the behavior of the amplification factor shown in FIG.~\ref{fig:PM_F}. A larger \(y\) corresponds to a smaller amplification factor, resulting in weaker lensing-induced modulation and thus smaller waveform discrepancies. Furthermore, when \(y = 100\), the difference between the lensed and unlensed waveforms falls below the characteristic sensitivity curve, indicating that the detector may not be able to effectively distinguish the lensed and unlensed waveforms in this regime.

\begin{figure}[h]
\centering
\includegraphics[width=8cm]{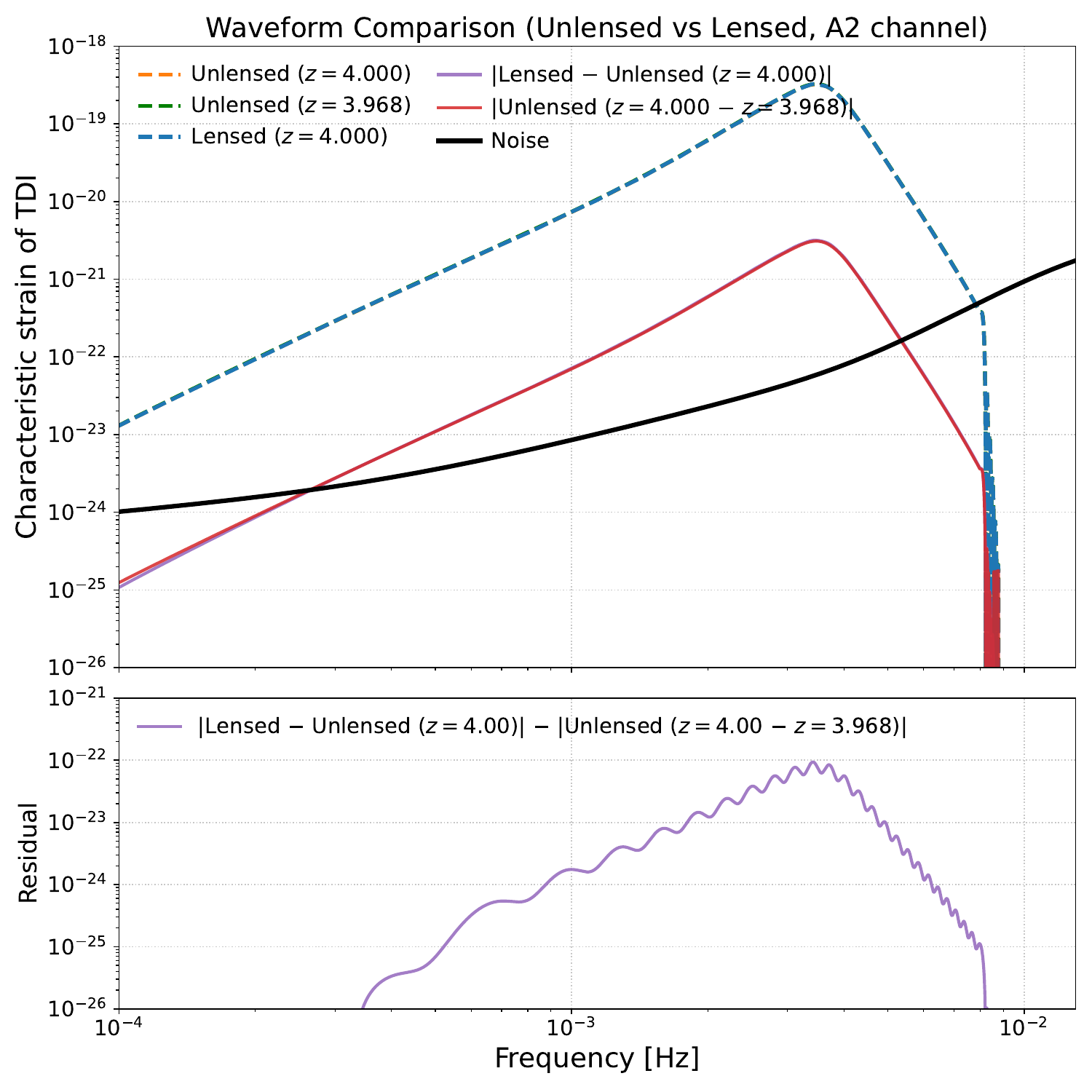}
\caption{Comparison of the characteristic strain spectra for lensed and unlensed gravitational wave signals in the A2 TDI channel. The top panel shows the lensed waveform ($z=4.000$, $M_{Lz}=3\times10^6,M_\odot$, $y=10$), the unlensed waveforms ($z=4.000$ and $z=3.968$), their differences, and the detector noise curve. The bottom panel presents the residuals, defined as the difference between the absolute lensed–unlensed discrepancy at $z=4.000$ and the corresponding unlensed–unlensed difference between $z=4.000$ and $z=3.968$.}
\label{fig:un_dl}
\end{figure}

We also analyzed whether varying only the luminosity distance could produce changes in the unlensed waveform similar to lensing effects. The binary masses in the observer frame were kept fixed ($\mathcal{M}c=2176343.4\ M_\odot$), and only the luminosity distance was adjusted. The lens parameters for the lensed waveform were a redshifted lens mass $M_{Lz} = 3 \times 10^6 \ M_\odot$ and an impact parameter $y = 10$. The waveform differences induced by lensing and luminosity distance variations are shown in the upper panel of FIG.~\ref{fig:un_dl}. We found that simply comparing the lensed and unlensed waveforms, or the differences between the lensed and unlensed waveforms and those induced by luminosity distance variations, does not allow a straightforward distinction between lensing effects and luminosity distance changes. Therefore, we further took the difference between these two sets of differences, as shown in the bottom panel of FIG.~\ref{fig:un_dl}. The results indicate that at low frequencies ($8\times10^{-3}$ Hz), the effects of lensing and luminosity distance variations are nearly identical; however, as the frequency increases, the differences between them gradually emerge. Moreover, the curve is no longer smooth because the magnification factor introduced by lensing oscillates with frequency, which is consistent with the result in Eq.~(\ref{eq:Fw}).

\subsection{3.2 Parameter estimation and model selection}

To investigate the effects of the binary's total mass, the source redshift, the redshifted lens mass, and the impact parameter y, we adopt a controlled-variable approach and analyze 9 lensed GW signals, as summarized in Table~\ref{tab:para}. In simulating lensed GW data, the coalescence time $t_c$ to the 149th day. In addition, $\phi_c$, $\lambda$, and $\beta$ denote the coalescence phase, the ecliptic longitude and latitude of the source, respectively, which have negligible influence on the ability to distinguish lensed GW waveforms. Therefore, in this study, we fix their values to $(\phi_c, \lambda, \beta) = (4.32, 4.25, -0.15)$.

Before performing parameter estimation, we first compute the mismatch ($\mathcal{MM}$) between the lensed and unlensed waveforms to assess whether the two models can be effectively distinguished. The definition of the $\mathcal{MM}$ is given in Eqs.~(\ref{eq:mis}, \ref{eq:inner}).
\begin{equation}
\mathcal{MM} = 1 - \langle h(\theta^S)|h^L(\theta^S,\theta^L)\rangle ,
\label{eq:mis}
\end{equation}
where $\langle.|.\rangle$ is
the noise-weighted inner product, defined as \cite{Usman2016CQGra}
\begin{equation}
\langle h(\theta^S)|h^L(\theta^S,\theta^L)\rangle\equiv4\mathrm{~}\Re\int_{f_{\mathrm{low}}}^{f_{\mathrm{high}}}\mathrm{d}f\mathrm{~}\frac{{h}^*(f;\theta^S){h}^L(f;\theta^S, \theta^L)}{S_{\mathrm{n}}(f)},
\label{eq:inner}
\end{equation}
where $*$ denotes complex conjugation. Note that only in the calculation of $\mathcal{MM}$, the normalized waveforms are used. The $\mathcal{MM}$ between the lensed and unlensed waveforms for each case is listed in the last column of Table.~\ref{tab:para}.

Next, we simulate lensed GW data and performed parameter estimation using both lensed and unlensed models. We compute the Bayes factor to assess whether the two scenarios can be distinguished. The inference was carried out using the \texttt{Bilby} \cite{Ashton2019ApJS} framework with the \texttt{nessai} \cite{Wu2025arXiv, williams2025, Williams2021PhRvD, Williams2023MLST} nested sampling algorithm.

\begin{sidewaystable*}[htbp]
\centering   
\caption{Parameter estimation results under different lensing configurations.
Each block corresponds to a lensed GW signal simulated with specific lensing parameters ($M_{Lz}$, $y$, $z_s$, and $M_s$) and its corresponding unlensed counterpart. $\mathcal{M}$ denotes the redshifted chirp mass, $d_L$ the luminosity distance, $t_c$ the coalescence time, and $\phi_c$ the coalescence phase, while $\lambda$ and $\beta$ represent the ecliptic longitude and latitude of the source in the sky. $y$ is the lens–source misalignment parameter, and $M_{Lz}$ is the redshifted lens mass. $\ln B^{L/N}$ is the natural logarithm of the Bayes factor comparing the lensed and unlensed models, SNR denotes the signal-to-noise ratio in Taiji for the difference between the lensed and unlensed waveforms, and $\mathcal{MM}$ represents the mismatch between the two waveforms. In this study, we fix $(t_c, \phi_c, \lambda, \beta) = (149 $\text{day}$, 4.32, 4.25, -0.15)$}.
\label{tab:para}
\renewcommand{\arraystretch}{1}
\resizebox{0.95\textwidth}{!}{ 
\begin{tabular}{|l|c|c|c|c|c|c|c|c|c|c|c|c|}
\hline
 & Model & $\mathcal{M}$ (M$_\odot$) & $d_L$ (Mpc) & $t_c (\text{day})$ & $\phi_c$ & $\lambda$ & $\beta$ & $y$ & $M_{Lz}$ (M$_\odot$) & $\ln B^{L/N}$ & SNR & $\mathcal{MM}$ \\
\hline
\multirow{2}{*}{\makecell[c]{1. $M_{Lz} = 3\times10^5$, $y=10$\\ $z_s=4$, $M_s=10^6$}} 
& Lensed   & $2176535^{+213}_{-223}$ & $36582^{+159}_{-170}$ & $149^{+1.8\times10^{-5}}_{-2\times10^{-5}}$ & $1.19^{+3.14}_{-0.01}$ & $4.26^{+0.01}_{-0.01}$ & $-0.15^{+0.01}_{-0.01}$ & $10.2^{+0.2}_{-0.2}$ & $290074^{+12363}_{-12032}$ & \multirow{2}{*}{1051} & \multirow{2}{*}{54} & \multirow{2}{*}{$3.61\times10^{-5}$} \\
\cline{2-10}
& Unlensed & $2175752^{+199}_{-188}$ & $36443^{+165}_{-172}$ & $149^{+1.9\times10^{-5}}_{-2.1\times10^{-5}}$ & $4.33^{+0.01}_{-3.14}$ & $4.27^{+0.01}_{-0.01}$ & $-0.15^{+0.01}_{-0.01}$ & -- & -- &  &  &  \\
\hline
\multirow{2}{*}{\makecell[c]{2. $M_{Lz} = 3\times10^6$, $y=10$\\ $z_s=4$, $M_s=10^6$}} 
& Lensed   & $2176508^{+194}_{-191}$ & $36586^{+170}_{-166}$ & $149^{+1.8\times10^{-5}}_{-2.0\times10^{-5}}$ & $4.32^{+0.01}_{-3.14}$ & $4.26^{+0.01}_{-0.01}$ & $-0.15^{+0.01}_{-0.01}$ & $10.1^{+0.2}_{-0.2}$ & $2918936^{+97539}_{-100143}$ & \multirow{2}{*}{1501} & \multirow{2}{*}{56} & \multirow{2}{*}{$4.80\times10^{-5}$} \\
\cline{2-10}
& Unlensed & $2176507^{+196}_{-197}$ & $36585^{+166}_{-173}$ & $149^{+1.9\times10^{-5}}_{-1.9\times10^{-5}}$ & $1.17^{+3.14}_{-0.01}$ & $4.26^{+0.01}_{-0.01}$ & $-0.14^{+0.01}_{-0.01}$ & -- & -- &  &  & \\
\hline
\multirow{2}{*}{\makecell[c]{3. $M_{Lz} = 3\times10^7$, $y=10$\\ $z_s=4$, $M_s=10^6$}} 
& Lensed   & $2176508^{+203}_{-204}$ & $36587^{+169}_{-166}$ & $149^{+1.8\times10^{-5}}_{-2.0\times10^{-5}}$ & $1.19^{+3.14}_{-0.01}$ & $4.26^{+0.01}_{-0.01}$ & $-0.14^{+0.01}_{-0.01}$ & $10.0^{+0.12}_{-0.10}$ & $30017892^{+609847}_{-627396}$ & \multirow{2}{*}{1581} & \multirow{2}{*}{57}  & \multirow{2}{*}{$4.81\times10^{-5}$} \\
\cline{2-10}
& Unlensed & $2176021^{+189}_{-198}$ & $36581^{+165}_{-165}$ & $149^{+1.8\times10^{-5}}_{-1.9\times10^{-5}}$ & $4.32^{+0.01}_{-3.14}$ & $4.26^{+0.01}_{-0.01}$ & $-0.15^{+0.01}_{-0.01}$ & -- & -- &  & & \\
\hline
\multirow{2}{*}{\makecell[c]{4. $M_{Lz} = 3\times10^6$, $y=30$\\ $z_s=4$, $M_s=10^6$}} 
& Lensed   & $2176506^{+187}_{-192}$ & $36588^{+158}_{-163}$ & $149^{+1.7\times10^{-5}}_{-1.9\times10^{-5}}$ & $4.32^{+0.01}_{-3.14}$ & $4.26^{+0.01}_{-0.01}$ & $-0.15^{+0.01}_{-0.01}$ & $28.7^{+4.7}_{-3.3}$ & $3269051^{+882686}_{-848161}$ & \multirow{2}{*}{17} & \multirow{2}{*}{6} & \multirow{2}{*}{$6.15\times10^{-7}$} \\
\cline{2-10}
& Unlensed & $2176510^{+191}_{-205}$ & $36587^{+160}_{-160}$ & $149^{+1.8\times10^{-5}}_{-2.0\times10^{-5}}$ & $1.19^{+3.14}_{-0.01}$ & $4.26^{+0.01}_{-0.01}$ & $-0.15^{+0.01}_{-0.01}$ & -- & -- &  & &  \\
\hline
\multirow{2}{*}{\makecell[c]{5. $M_{Lz} = 3\times10^6$, $y=50$\\ $z_s=4$, $M_s=10^6$}} 
& Lensed   & $2176510^{+192}_{-194}$ & $36586^{+159}_{-160}$ & $149^{+1.7\times10^{-5}}_{-1.9\times10^{-5}}$ & $1.19^{+3.14}_{-0.01}$ & $4.26^{+0.01}_{-0.01}$ & $-0.15^{+0.01}_{-0.01}$ & $164.7^{+106}_{-114}$ & $14459019^{+14801673}_{-13734155}$ & \multirow{2}{*}{-1} & \multirow{2}{*}{2} & \multirow{2}{*}{$7.99\times10^{-8}$}  \\
\cline{2-10}
& Unlensed & $2176507^{+201}_{-193}$ & $36583^{+168}_{-159}$ & $149^{+1.8\times10^{-5}}_{-1.9\times10^{-5}}$ & $1.19^{+3.14}_{-0.01}$ & $4.26^{+0.01}_{-0.01}$ & $-0.14^{+0.01}_{-0.01}$ & -- & -- &  & & \\
\hline
\multirow{2}{*}{\makecell[c]{6. $M_{Lz} = 3\times10^6$, $y=10$\\ $z_s=5$, $M_s=10^6$}} 
& Lensed   & $2611646^{+336}_{-322}$ & $47691^{+323}_{-322}$ & $149^{+2.7\times10^{-5}}_{-3.0\times10^{-5}}$ & $1.19^{+3.14}_{-0.01}$ & $4.26^{+0.02}_{-0.02}$ & $-0.15^{+0.01}_{-0.01}$ & $10.0^{+0.25}_{-0.22}$ & $2990268^{+126635}_{-132903}$ & \multirow{2}{*}{857} & \multirow{2}{*}{41} & \multirow{2}{*}{$4.80\times10^{-5}$}\\
\cline{2-10}
& Unlensed & $2611628^{+347}_{-332}$ & $47689^{+320}_{-325}$ & $149^{+2.7\times10^{-5}}_{-3.1\times10^{-5}}$ & $1.19^{+3.14}_{-0.01}$ & $4.26^{+0.02}_{-0.02}$ & $-0.15^{+0.01}_{-0.02}$ & -- & -- &  & &  \\
\hline
\multirow{2}{*}{\makecell[c]{7. $M_{Lz} = 3\times10^6$, $y=10$\\ $z_s=6$, $M_s=10^6$}} 
& Lensed   & $3046765^{+530}_{-521}$ & $59251^{+606}_{-569}$ & $149^{+3.7\times10^{-5}}_{-4.7\times10^{-5}}$ & $1.19^{+3.14}_{-0.02}$ & $4.25^{+0.02}_{-0.02}$ & $-0.16^{+0.02}_{-0.02}$ & $9.9^{+0.3}_{-0.3}$ & $3067849^{+178360}_{-180718}$ & \multirow{2}{*}{519} & \multirow{2}{*}{32} & \multirow{2}{*}{$4.79\times10^{-5}$} \\
\cline{2-10}
& Unlensed & $3046740^{+523}_{-524}$ & $59287^{+592}_{-554}$ & $149^{+3.9\times10^{-5}}_{-4.6\times10^{-5}}$ & $1.20^{+3.14}_{-0.02}$ & $4.25^{+0.02}_{-0.02}$ & $-0.16^{+0.02}_{-0.02}$ & -- & -- &  & & \\
\hline
\multirow{2}{*}{\makecell[c]{8. $M_{Lz} = 3\times10^6$, $y=10$\\ $z_s=4$, $M_s=10^5$}} 
& Lensed   & $217634^{+2.9}_{-2.7}$ & $36596^{+262}_{-254}$ & $149^{+2.7\times10^{-5}}_{-3.0\times10^{-5}}$ & $1.19^{+3.14}_{-0.01}$ & $4.25^{+0.01}_{-0.01}$ & $-0.14^{+0.01}_{-0.01}$ & $10.1^{+0.85}_{-0.69}$ & $2932564^{+407474}_{-404911}$ & \multirow{2}{*}{78} & \multirow{2}{*}{13} & \multirow{2}{*}{$4.80\times10^{-5}$} \\
\cline{2-10}
& Unlensed & $217631^{+2.9}_{-2.8}$ & $36642^{+276}_{-283}$ & $149^{+3.0\times10^{-5}}_{-3.3\times10^{-5}}$ & $4.32^{+0.01}_{-3.14}$ & $4.25^{+0.02}_{-0.02}$ & $-0.14^{+0.02}_{-0.02}$ & -- & -- &  & & \\
\hline
\multirow{2}{*}{\makecell[c]{9. $M_{Lz} = 3\times10^6$, $y=10$\\ $z_s=4$, $M_s=10^7$}} 
& Lensed   & $21765749^{+11004}_{-11648}$ & $36373^{+1521}_{-979}$ & $149^{+1.4\times10^{-4}}_{-2.7\times10^{-4}}$ & $1.22^{+3.13}_{-0.29}$ & $4.28^{+0.06}_{-0.74}$ & $-0.17^{+0.05}_{-0.09}$ & $9.9^{+1.4}_{-1.0}$ & $3082607^{+710795}_{-693803}$ & \multirow{2}{*}{30} & \multirow{2}{*}{12} & \multirow{2}{*}{$3.16\times10^{-5}$} \\
\cline{2-10}
& Unlensed & $21764181^{+9794}_{-10327}$ & $36233^{+1464}_{-890}$ & $149^{+1.3\times10^{-4}}_{-1.9\times10^{-4}}$ & $4.08^{+0.29}_{-2.90}$ & $4.29^{+0.05}_{-0.69}$ & $-0.16^{+0.05}_{-0.09}$ & -- & -- &  & &  \\
\hline
\end{tabular}
}
\end{sidewaystable*}

Due to the long observation duration of space-based GW detectors and the wide prior ranges for the parameters of MBHB, performing parameter estimation via nested sampling directly over such broad priors would consume substantial computational resources and be inefficient. In this study, we first use the Fisher matrix to compute the parameter uncertainties~\cite{Finn1992PhysRevD,Vallisneri2008PhysRevD}, and then limit the prior ranges of $\{\mathcal{M}_c, t_c, d_L, y_L, M_{L, z}\}$ to ten times these uncertainties before carrying out the parameter estimation. The limited priors are not applied to parameters $\{\varphi_c, \lambda, \beta\}$ as their posterior distributions may exhibit multimodality~\cite{Marsat2021PhRvD}. Since the work of this paper aims to investigate the distinguishability of lensed GWs, this approach would significantly improve computational efficiency without compromising the accuracy of the results.

\begin{figure*}[htbp]
\centering
\subfigure{
    \includegraphics[width=0.62\textwidth]{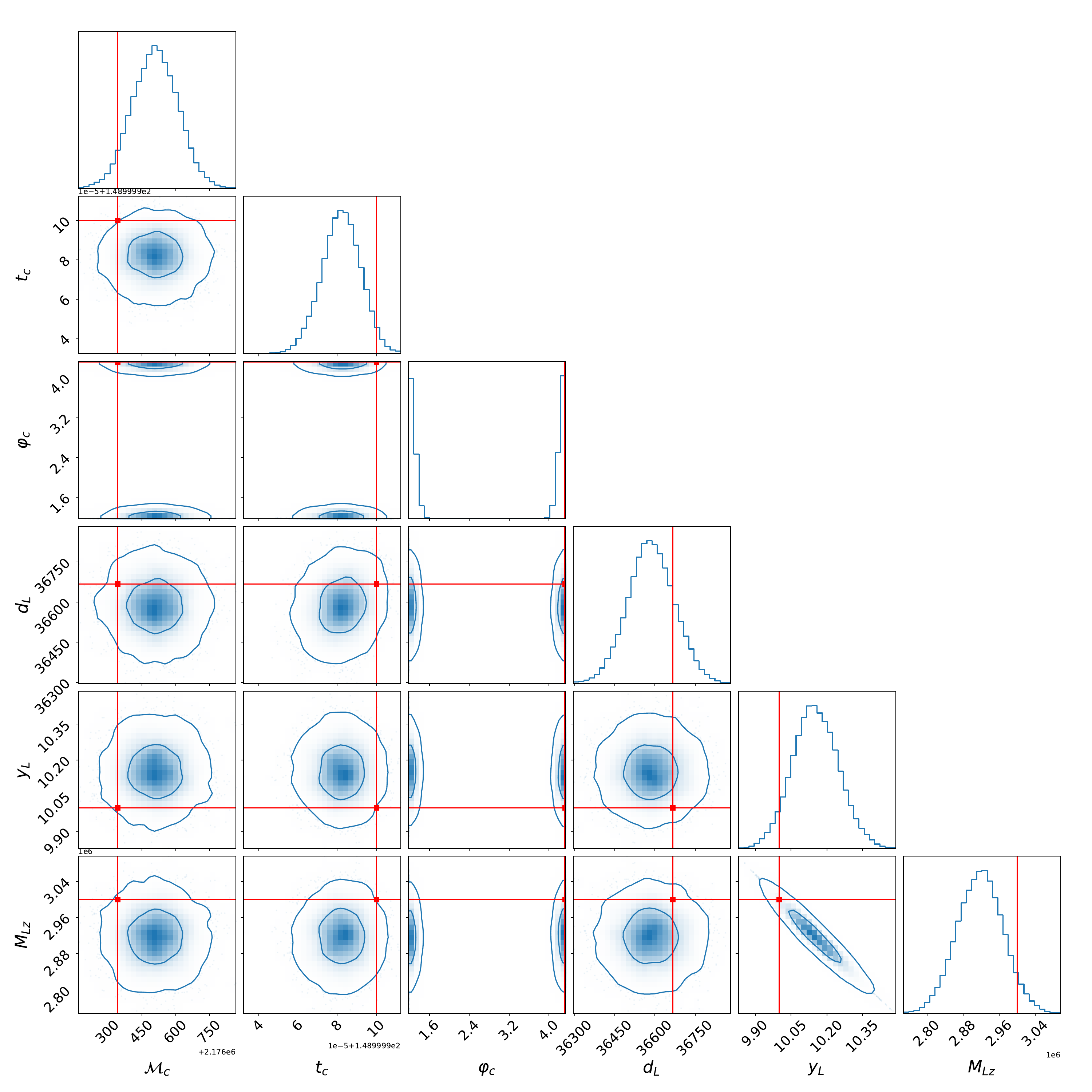}
}
\subfigure{
    \includegraphics[width=0.62\textwidth]{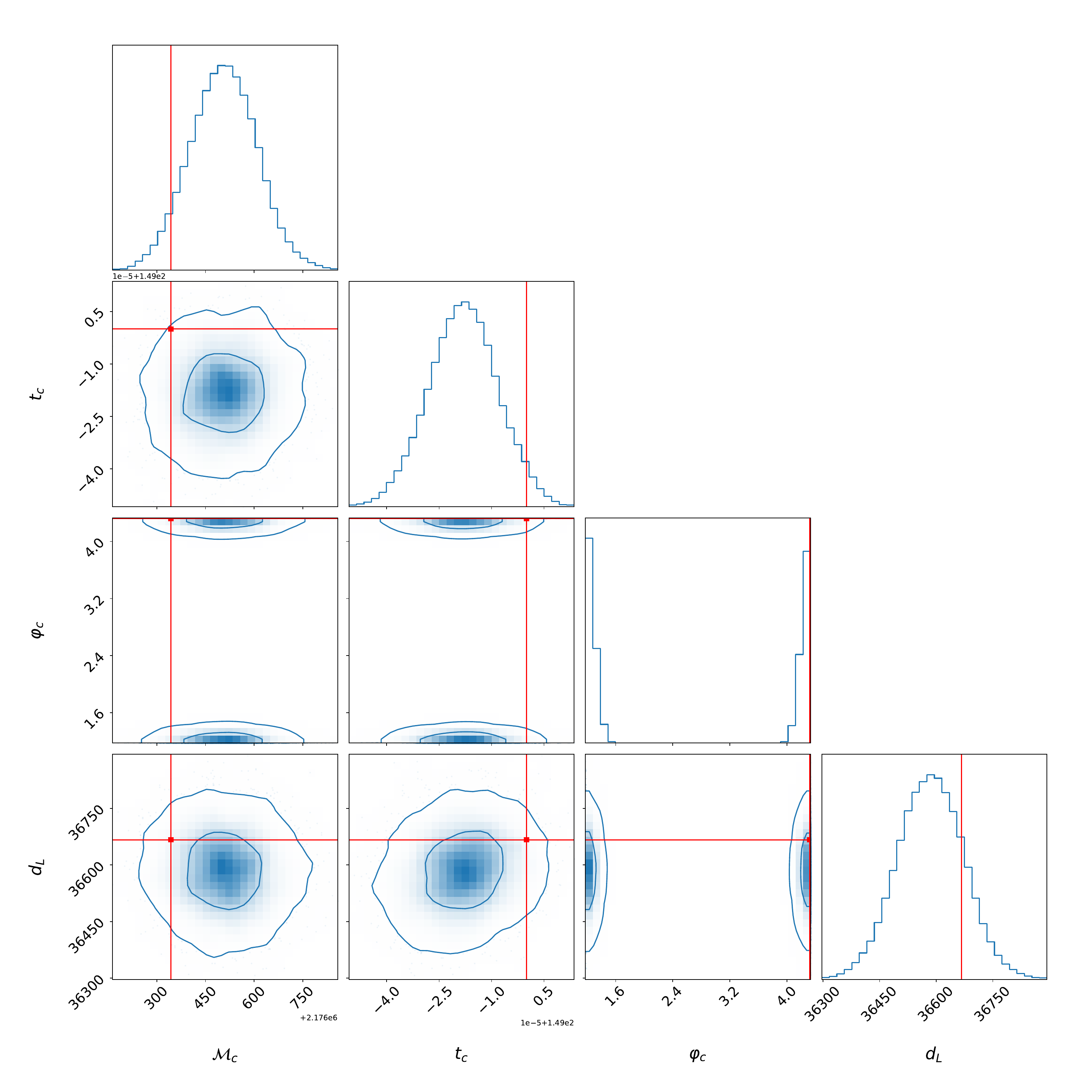}
}
\caption{Posterior distributions of the source and lens parameters for the lensed (top) and unlensed (bottom) models corresponding to Case 2 in Table~\ref{tab:para}. The vertical and horizontal red lines indicate the injected values. The Bayes factor between the lensed and unlensed hypotheses is $\ln B^{L/N} = 1501$. The sampling frequency of the data is 0.03.}
\label{fig:case2}
\end{figure*}

\begin{figure*}[htbp]
\centering
\subfigure{
    \includegraphics[width=0.62\textwidth]{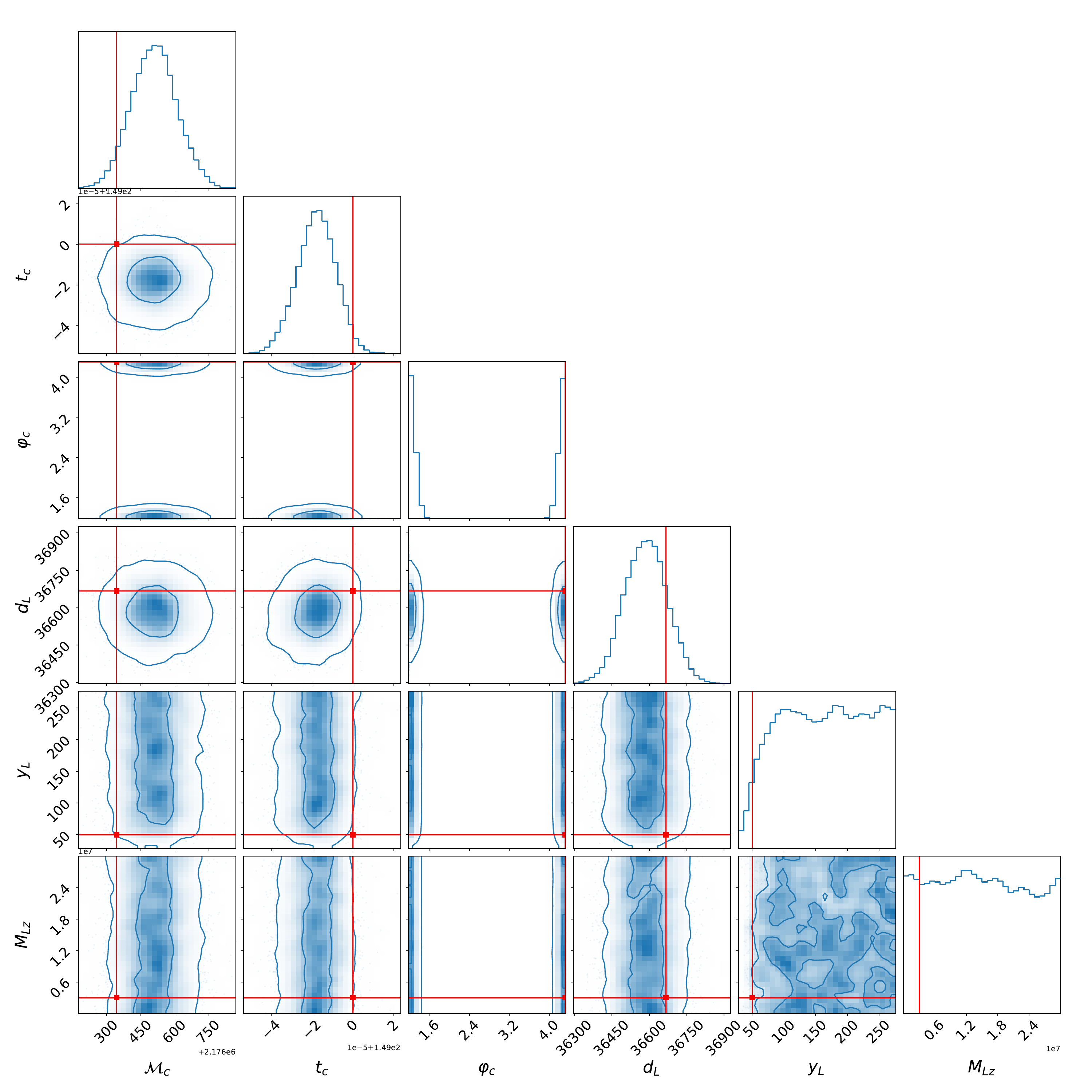}
}
\subfigure{
    \includegraphics[width=0.62\textwidth]{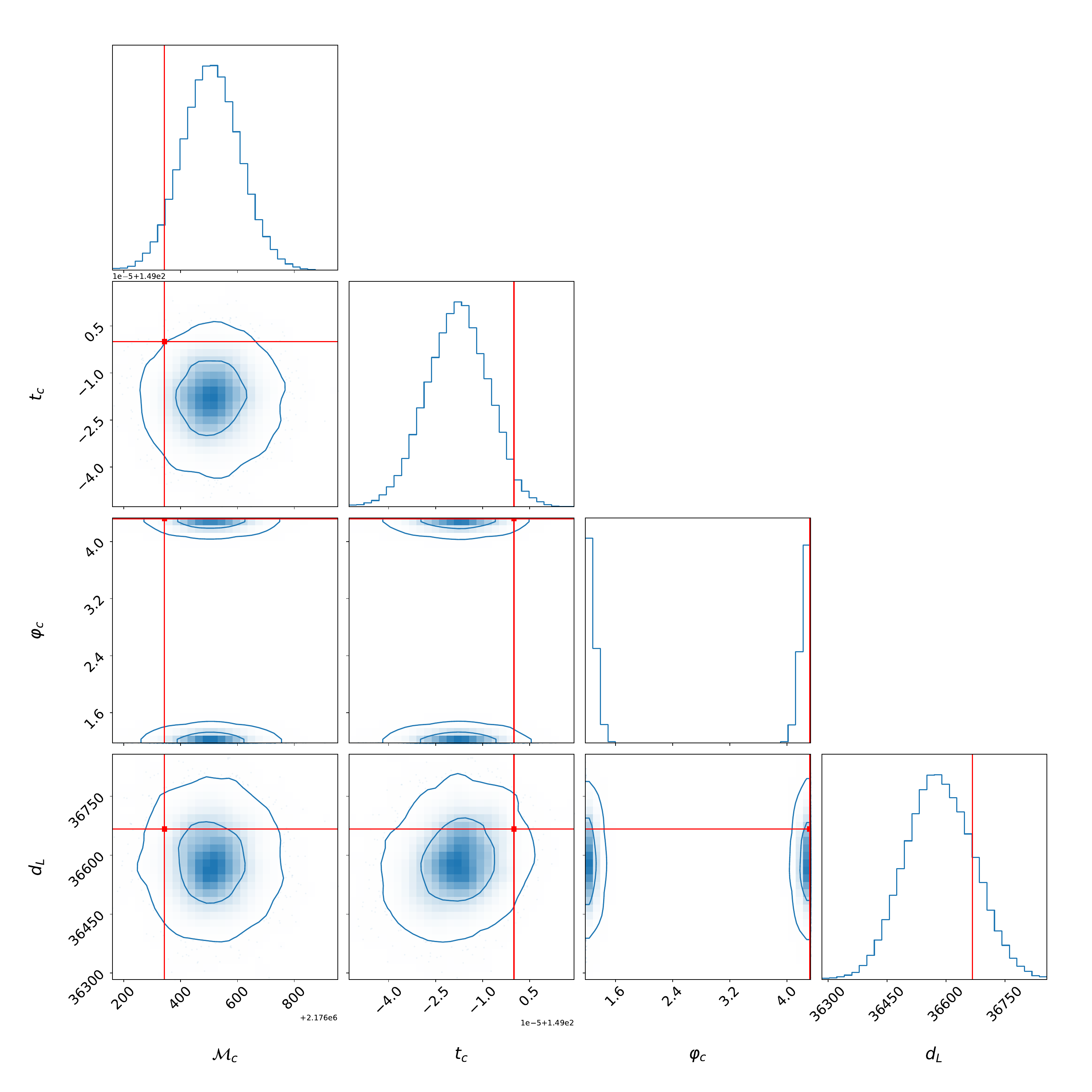}
}
\caption{Similar to Figure~\ref{fig:case2}, but presenting the parameter estimation results for case 5, with a Bayes factor of $\ln B^{L/N} = -1$.}
\label{fig:case4}
\end{figure*}

\begin{figure*}[htbp]
\centering
    \includegraphics[width=0.95\textwidth]{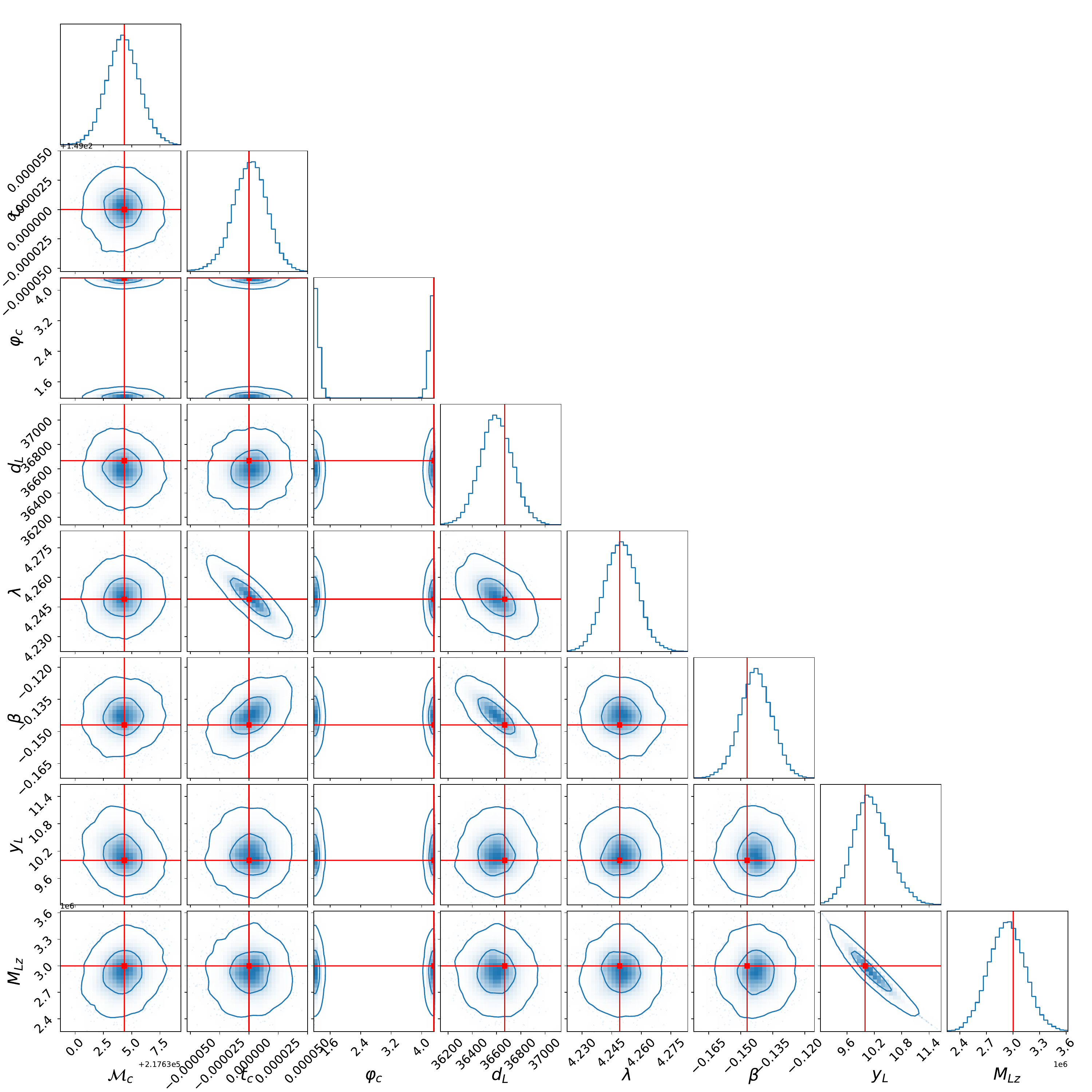}
\caption{Parameter estimation results for the lens model in Case 8, presenting the complete set of inferred lensing parameters. The sampling frequency of the data is 0.3.}
\label{fig:case8}
\end{figure*}

We select two dataset (Case 2 and Case 5) from Table~\ref{tab:para} to illustrate the parameter estimation results obtained with both the lensed and unlensed models. The corresponding results are presented in FIG.~\ref{fig:case2} and FIG.~\ref{fig:case4}. Since the sky location parameters are not among the key factors affecting the conclusions of this work, their posterior distributions are not shown in FIG.~\ref{fig:case2} and FIG.~\ref{fig:case4}. Nevertheless, these parameters are still included in our parameter estimation procedure. For completeness, FIG.~\ref{fig:case8} presents an example of the posterior distributions when all parameters are estimated.

From the results shown in FIG.~\ref{fig:case2}, we observe that the parameter estimation outcomes for both the lensed and unlensed models lie within the $95\%$ credible intervals. However, a comparison of the Bayes factors indicates a stronger preference for the lensed model. Furthermore, FIG.~\ref{fig:case2} demonstrates that the degeneracy between the chirp mass and the luminosity distance is effectively broken. This is because the mass not only affects the amplitude of the waveform but also modifies the evolution of its frequency, enabling the two parameters to be distinguished. The degeneracy between the luminosity distance and the lensing parameters $M_{Lz}$ and $y$ is also broken, as the lensing magnification factor depends on frequency, with different frequencies corresponding to different magnification factors, thereby allowing these parameters to be distinguished (see FIG.~\ref{fig:channel} and FIG.~\ref{fig:un_dl}). On the other hand, a significant degeneracy remains between $M_{Lz}$ and $y$, which could potentially be broken by incorporating electromagnetic counterparts, and we plan to explore this approach in future work.

From the results shown in FIG.~\ref{fig:case4}, we observe that the lensed and unlensed models yield consistent estimates for the unlensed parameters ($\mathcal{M}_c,d_L, t_c, \phi_c$), with all results lying within the 95\% credible intervals. However, in the lensed model, the posterior distributions of the lensing parameters ($M_{Lz}$ and $y$) almost coincide with their priors, indicating that these parameters are not effectively constrained. Furthermore, both models give a log Bayes factor of $\ln B^{L/N} = -1$, which is closely related to the fact that the SNR of the difference between the lensed and unlensed waveforms is only 2. Such a small difference makes it statistically impossible to distinguish between the two models, thereby preventing effective constraints on the lensing parameters.

Compared with FIG.~\ref{fig:case2} and  FIG.~\ref{fig:case4}, the result shown in  FIG.~\ref{fig:case8} corresponds to a sampling frequency of 0.3, which is ten times higher. As confirmed by the results in Table~\ref{tab:para} and the waveform plots, the parameter constraints obtained in FIG.~\ref{fig:case8} are also more precise.

We now analyze the results summarized in Table~\ref{tab:para}, focusing on the influence of various physical parameters on parameter estimation and model selection. First, to isolate the effect of lens mass (Cases 1, 2, and 3), we vary only the redshifted lens mass $M_{Lz}$ while keeping all other parameters fixed. The results show that the SNR of the difference between lensed and unlensed waveforms exceeds 50, leading to parameter estimation outcomes that strongly favor the lensed model. Moreover, the Bayes factor increases with redshifted lens mass, as a more massive lens induces stronger modulation through the magnification factor, resulting in more pronounced waveform oscillations and larger differences between the two models.
Second, we examine the effect of the impact parameter $y$ (Cases 2, 4, and 5), with $y=10$, 30, and 50. In Case 5, the lensed and unlensed models cannot be distinguished, as the SNR of the waveform difference is only 2, yielding parameter estimation results that fail to discriminate between the models. Accordingly, cases with $y\ge50$ are not considered further. Comparing Cases 2, 4, and 5, smaller $y$ values produce larger magnification factors, enhancing the waveform differences and making the models easier to distinguish. This trend is consistent with the Bayes factor. Our analysis indicates that when the SNR of the waveform difference exceeds 5, the two models can be effectively distinguished (corresponding to $\ln B^{L/N}>5$).
Third, we investigate the effect of the source luminosity distance (Cases 2, 6, and 7), corresponding to redshifts $z=4$, 5, and 6. As redshift increases, the SNR of the waveform difference decreases, reflecting the inverse scaling of GW amplitude with luminosity distance. Nevertheless, in all these cases, the SNR remains well above 5, ensuring that the lensed and unlensed models can still be effectively distinguished.
Finally, we consider the impact of the total mass of the binary system in the source frame (Cases 2, 8, and 9). Case 2 produces the largest waveform difference SNR, consistent with the top panel of FIG.~\ref{fig:channel}, and consequently the largest $\ln B^{L/N}$, making it the easiest case for distinguishing the two models. For Cases 8 and 9, although the SNRs are similar, Case 8 yields a higher $\ln B^{L/N}$ because the binary components in Case 8 have lower masses, which generate higher-frequency GW signals, as shown in FIG.~\ref{fig:channel}. These higher-frequency signals improve the parameter estimation accuracy, resulting in a larger $\ln B^{L/N}$.
In summary, both lens mass and impact parameter play decisive roles in determining the distinguishability of lensed and unlensed models, whereas luminosity distance primarily affects the absolute SNR without hindering model selection. Higher sampling frequency further enhances the effectiveness of Bayesian inference. Overall, these results suggest that strong lensing signatures in GW data can be robustly identified across a wide range of astrophysical conditions, provided that the waveform difference achieves an SNR $>$ 5. Across Table~\ref{tab:para}, the mismatch $\mathcal{MM}$ quantifies waveform differences and directly determines the differential SNR. Since SNR is closely correlated with the Bayes factor $\ln B^{L/N}$, larger mismatches correspond to higher SNR and stronger support for the lensed model, whereas very small mismatches render lensing effects nearly undetectable.

\section{4. Conclusion}
Previous studies suggest that once space-based GW detectors such as LISA, Taiji and TianQin are operational, there will be a non-negligible probability of observing lensed mergers of MBHB \cite{Pau2017arXiv, Hu2017NSRev, Luo2016CQGra}. In this work, we investigate the lensing signals of MBHB in the Taiji frequency band, focusing on the wave-optics effects under the PM lens model. We simulate 90 days of GW data and, using a controlled variable approach, analyze how variations in lens mass, impact parameter $y$, luminosity distance, and the total mass of the binary system affect the parameter estimation accuracy under the lens model. We also compute the Bayes factor to compare the lensed and unlensed scenarios.

In this study, we systematically analyzed 9 representative cases, and the posterior distributions of the parameters under the two models are summarized in Table~\ref{tab:para}. Prior to parameter estimation, we computed the $\mathcal{MM}$ between the lensed and unlensed waveforms to assess whether the two models could be effectively distinguished. Our results indicate that an $\mathcal{MM}$ value larger than $10^{-7}$ generally corresponds to a waveform difference SNR exceeding 5, which is sufficient to discriminate between the lensed and unlensed models.
A larger redshifted lens mass increases the magnification factor, thereby enhancing both the waveform difference and the corresponding SNR, which facilitates distinguishing the lensed model from the unlensed one. Conversely, a larger impact parameter $y$ reduces the magnification factor and weakens the waveform difference, resulting in a lower SNR and making the two models harder to separate. In our analysis, for relatively large values of $y$, the lensing effect becomes significantly weaker, and the waveform difference and corresponding SNR decrease to levels where the two models can no longer be effectively distinguished. Specifically, under the parameter settings considered in this study, the models become difficult to distinguish when $y \gtrsim 50$.
Regarding the source luminosity distance, as the distance increases, both the waveform difference and the corresponding SNR decrease, leading to a reduction in the Bayes factor. Nevertheless, for redshifts up to $z \leq 6$, the waveform difference SNR remains above the threshold, allowing effective discrimination between the two models.
The total mass of the binary system also affects parameter estimation, though the relationship is not monotonic. In general, effective distinction between the lensed and unlensed models is achieved when the GW frequencies generated by the binary system substantially overlap with the detector’s sensitivity band and when the waveform difference SNR is sufficiently high (exceeding 5). In Case 8, the improvement in parameter estimation accuracy is mainly due to the merger of lower-mass MBHB, which produces higher-frequency GW signals that better fall within the detector’s sensitive range, thereby significantly enhancing the estimation precision, albeit at the cost of increased computational demand.
From the posterior distributions, we observe that the lensed parameters $\theta^L$ and the unlensed parameters $\theta^S$ do not exhibit significant degeneracy, although a partial degeneracy exists between the lens mass and the impact parameter within the lensed parameters. Except for Case 5, where the posterior largely reflects the prior, the parameter estimation accuracy in all other cases falls within the 95\% credible intervals.
Compared to ground-based GW detectors, space-based detectors can observe signals from massive black hole mergers with SNRs up to several thousand and over much longer durations. This enables the detection of lensing events with larger impact parameters $y$, providing greater opportunities to probe the properties of the lensing objects.

Finally, it should be noted that the lensing signal model and the simulated data used in this study are based on idealized assumptions. For instance, we assume that the observed data contain only the GW signals from massive binary black holes and consider only the case of a single lens along the line of sight, neglecting the possibility of multiple lenses. In actual Taiji observations, GW signals from other sources, such as double white dwarfs, would also be present in the data, which is not accounted for in this study. In future work, we plan to extend our analysis to lensing signals in GW data that more closely reflect realistic observational conditions.

\section{Acknowledge}
We thank Huan Zhou and Qi-Yuan Yang for useful discussion.
This work is supported by National Key Research and Development Program of China Grant No. 2024YFC2207300  and No. 2021YFC2201903. Peng Xu is supported by the International Partnership Program of the Chinese Academy of Sciences, Grant No.
025GJHZ2023106GC.

\nocite{*}

\bibliographystyle{apsrev4-2} 
\bibliography{apssamp}

\section{APPENDIX: Simulation of Gravitational-Wave Signals from Massive Binary Black Hole Mergers Based on Taiji}

Here we provide the detailed procedure for simulating the GW signals used in the main text. Following \cite{Wahlquist1987GReGr, Vallisneri2005PhRvD}, the relative frequency shift of laser link $ij$ (i.e. spacecraft$_j$ $\rightarrow$ spacecraft$_i$, with $i, j \in \{1, 2, 3\}$) induced by GW reads 

\begin{equation}
\begin{aligned}
y_{ij}(t) &\equiv 
   \frac{\nu_{\rm receive} - \nu_{\rm send}}{\nu_{\rm send}} \\[6pt]
&\approx 
   \frac{\bm{\hat{n}}_{ij}(t) \otimes \bm{\hat{n}}_{ij}(t)}
        {2\left(1-\bm{\hat{k}} \cdot \bm{\hat{n}}_{ij}(t)\right)} \\
&\quad : \Bigg[
   \bm{h}\!\left(t - d_{ij}(t) - \frac{\bm{\hat{k}} \cdot \bm{R}_j(t)}{c}\right) \\
&\qquad -\;
   \bm{h}\!\left(t - \frac{\bm{\hat{k}} \cdot \bm{R}_i(t)}{c}\right)
   \Bigg].
\end{aligned}
\label{eq:yij}
\end{equation}
where $\bm{h}(t)$ is the GW tensor, $\hat{\bm{k}}$ is the unit vector along the direction of GW propagation, $\hat{\bm{n}}_{ij}(t)$ is the unit vector along laser link $ij$, $d_{ij}$ is the corresponding laser propagation time, : is contraction symbol, and $\bm{R}_{i, j}(t)$ represent the positions of spacecraft$_i$ and spacecraft$_j$, respectively. For later convenience, we further express $\bm{h}(t)$ in a mode-decomposition manner: 
\begin{equation}
\begin{aligned}
\bm{h}(t) &= \sum_{\alpha} \sum_{\ell m} 
   h_{\ell m}(t)\, K^{\ell m}_\alpha(\iota, \varphi_c)\, 
   \bm{e}^\alpha(\lambda, \beta, \psi), \\[6pt]
&\quad \alpha \in \{+, \times\}, \ \ 
   \ell m \in {\rm harmonics}.
\end{aligned}
\end{equation}
where $\bm{e}^\alpha (\lambda, \beta, \psi)$ is the polarization basis in the source frame of GW, with $\lambda, \beta, \psi$ being the source’s Ecliptic longitude, Ecliptic latitude and polarization angle, respectively. $K^{\ell m}_\alpha(\iota, \varphi_c)$ is constructed from the spin-weighted spherical harmonics, which depends on 2 parameters: the inclination angle $\iota$ and the reference (here we set it to the phase at coalescence $\varphi_c$). 

In realistic detection, the single-link measurements are dominated by laser frequency noise, whose amplitude is orders of magnitudes larger than GWs. Therefore, both LISA and Taiji require the implementation of time-delay interferometry (TDI) technology to effectively mitigate laser frequency noise through properly delaying and combining the raw measurements \cite{Tinto1999PhRvD, Ni1997SPIE, Tinto2014LRR, Armstrong1999ApJ}. Various TDI schemes have been proposed in the history \cite{Tinto2021LRR}, and for the non-rigid and rotating constellations of LISA and Taiji,  the 2nd-generation TDI should be employed \cite{Estabrook2000PhRvD, Tinto2004PhRvD}. By denoting $\textbf{D}_{i_1i_2}f(t) \equiv f[t - d_{i_1i_2}(t)]$ and $\textbf{D}_{i_1i_2i_3 ...} f(t) \equiv \textbf{D}_{i_1i_2}\textbf{D}_{i_2i_3}...f(t)$, the widely adopted Michelson-$X_2$ TDI variable is formulated as 
\begin{align}
X_2(t) = &\ \Big( 1 - {\textbf{D}}_{131} - {\textbf{D}}_{13121} + {\textbf{D}}_{1213131} \Big) 
            y_{12}(t) \nonumber \\ 
         &\ + \Big( 1 - {\textbf{D}}_{131} - {\textbf{D}}_{13121} + {\textbf{D}}_{1213131} \Big) 
            {\textbf{D}}_{12} y_{21}(t) \nonumber \\ 
         &\ - \Big( 1 - {\textbf{D}}_{121} - {\textbf{D}}_{12131} + {\textbf{D}}_{1312121} \Big) y_{13}(t) \nonumber \\
         &\ - \Big( 1 - {\textbf{D}}_{121} - {\textbf{D}}_{12131} + {\textbf{D}}_{1312121} \Big) 
            {\textbf{D}}_{13} y_{31}(t).
\label{eq:tdi}
\end{align}
The definitions for Michelson-$Y_2$ and Michelson-$Z_2$ can be obtained by cyclically permuting the indices with rule: $1 \rightarrow 2, 2 \rightarrow 3, 3 \rightarrow 1$. For Bayesian analysis, it is more convenient to use the quasi-noise-orthogonal combinations $A_2, E_2$ and $T_2$ \cite{Prince2002PhRvD}, defined as
\begin{equation}
\begin{aligned}
A_2 &\equiv \frac{Z_2 - X_2}{\sqrt{2}}, \quad 
E_2 \equiv \frac{X_2 - 2Y_2 + Z_2}{\sqrt{6}}, \\ 
T_2 &\equiv \frac{X_2 + Y_2 + Z_2}{\sqrt{3}}.
\end{aligned}
\label{eq:AET}
\end{equation}
The term ``orthogonal'' implies that when computing the likelihood function, the contributions of these channels can be directly multiplied.
In the low-frequency regime the response of $T_2$ channel is largely suppressed, thus only the “signal” channels $A_2, E_2$ will be considered in this study. 

To facilitate the calculation of statistics like the inner product, we then apply Fourier transform to the TDI response described above. For an arbitrary frequency-domain waveform $\tilde{h}_{\ell m}(f)$ that admits a decomposition of the form
\begin{equation}
\tilde{h}_{\ell m}(f) = \sum_{\ell m} \mathcal{A}_{\ell m}(f) e^{-i \Phi_{\ell m}(f)},
\label{eq:hlm}
\end{equation}
following the methodology of \cite{Marsat2021PhRvD}, we obtain 
\begin{equation}
\tilde{X}_2(f) = \sum_{\ell m} \mathcal{G}^{\ell m}_{X_2}(f, t_{f, \ell m}) \mathcal{A}_{\ell m}(f) e^{-i \Phi_{\ell m}(f)},
\label{eq:X2}
\end{equation}
where the time-frequency relationship for the $\ell m$ mode is 
\begin{equation}
t_{f, \ell m} \equiv  -\frac{1}{2\pi}\frac{d \Phi_{\ell m}}{df},
\end{equation}
and $X_2$ channel transfer function  $\mathcal{G}_{X_2}^{\ell m}(f, t_{f, \ell m})$ takes the form 
\begin{align}
&\mathcal{G}^{\ell m}_{X_2}(f, t_{f , \ell m}) \\
= &\ \Big( 1 - \tilde{\textbf{D}}_{131} - \tilde{\textbf{D}}_{13121} + \tilde{\textbf{D}}_{1213131} \Big) 
    \mathcal{G}_{12}^{\ell m}(f, t_{f, \ell m}) \nonumber \\
  &\ + \Big( 1 - \tilde{\textbf{D}}_{131} - \tilde{\textbf{D}}_{13121} + \tilde{\textbf{D}}_{1213131} \Big) 
     \tilde{\textbf{D}}_{12} \mathcal{G}^{\ell m}_{21}(f, t_{f, \ell m}) \nonumber \\
  &\ - \Big( 1 - \tilde{\textbf{D}}_{121} - \tilde{\textbf{D}}_{12131} + \tilde{\textbf{D}}_{1312121} \Big) 
    \mathcal{G}^{\ell m}_{13}(f, t_{f, \ell m}) \nonumber \\
  &\ - \Big( 1 - \tilde{\textbf{D}}_{121} - \tilde{\textbf{D}}_{12131} + \tilde{\textbf{D}}_{1312121} \Big) 
     \tilde{\textbf{D}}_{13} \mathcal{G}^{\ell m}_{31}(f, t_{f, \ell m}).
\label{eq:G_X2}
\end{align}
with $\tilde{\textbf{D}}_{i_1i_2i_3...}(f, t_{f, \ell m}) \equiv e^{i 2\pi f d_{i_1i_2i_3...}(t_{f, \ell m})}$. The single-link transfer function $
\mathcal{G}_{ij}^{\ell m}(f, t_{f, \ell m})$ is given by: 
\begin{align}
\mathcal{G}_{ij}^{\ell m}(f, t_{f, \ell m}) 
&= i \pi f d_{ij}(t_f) \, 
   {\rm sinc} \Big[ \pi f d_{ij}(t_f) \big(1-\bm{\hat{k}} \cdot \bm{\hat{n}}_{ij}(t_f)\big) \Big] \nonumber \\ 
&\quad \times e^{i \pi f \left[d_{ij}(t_f) + \frac{\bm{\hat{k}} \cdot (\bm{R}_i(t_f) + \bm{R}_j(t_f))}{c}\right]} F_{ij}^{\ell m}(t_f),
\end{align}
where  
\begin{equation}
F^{\ell m}_{ij}(t_f) \equiv \bm{\hat{n}}_{ij}(t_f) \otimes \bm{\hat{n}}_{ij}(t_f) : \sum_{\alpha} K^{\ell m }_\alpha(\iota, \varphi_c) \bm{e}^\alpha(\lambda, \beta, \psi).
\end{equation}
The responses of TDI-$A_2, E_2$ channels can be simply obtained according to their definitions. 
In summary, the frequency-domain responses of the TDI-$A_2, E_2$ channels to GWs can be expressed in a concise form \cite{Marsat2021PhRvD}:
\begin{equation}
h_{\tilde{A}_2,\tilde{E}_2}(f; \theta^S) = \sum_{\ell m} \mathcal{T}_{A_2, E_2}^{\ell m}(f, t_{f, \ell m}) \tilde{h}_{\ell m}(f),
\end{equation}
with $\mathcal{T}^{\ell m}_{A_2, E_2}$ being the transfer function. 
In this study, we employ the IMRPhenomD waveform implemented in the \texttt{WF4Py} package~\cite{Iacovelli:2022bbs,Iacovelli:2022mbg}, which describes the dominate $\ell = 2, |m| = 2$ modes of aligned-spin binary systems.

The instrumental noises are generated as Gaussian noises according to the noise model of Taiji. 
The noise budgets of both LISA and Taiji comprise 2 primary components: the Optical Metrology System (OMS) noise and the test-mass residual ACCeleration (ACC) noise. For the Taiji mission, the amplitude requirements for these noises are specified as $A_{\rm OMS} = 8 \times 10^{-12} \ \mathrm{m}/\sqrt{\mathrm{Hz}}$ and $A_{\rm ACC} = 3 \times 10^{-15} \ \mathrm{m}/\mathrm{s}^2/\sqrt{\mathrm{Hz}}$, respectively \cite{Luo2021PTEP}. Their respective power spectral densities (PSDs) are given by
\begin{equation}
P_{\rm OMS}(f) = A_{\rm OMS}^2 \left(\frac{2\pi f}{c}\right)^2 \left[1 + \left(\frac{2 \ {\rm mHz}}{f}\right)^4\right],
\label{eq:OMS}
\end{equation}

\begin{equation}
\begin{aligned}
P_{\rm ACC}(f) =\;& A_{\rm ACC}^2 \left(\frac{1}{2\pi f c}\right)^2 
   \left[1 + \left(\frac{0.4\,{\rm mHz}}{f}\right)^2\right] \\
 &\times \left[1 + \left(\frac{f}{8\,{\rm mHz}}\right)^4\right].
\end{aligned}
\label{eq:ACC}
\end{equation}

In above equations, the $(2\pi f /c )^2$ and  $[1 / (2\pi f c)]^2$ terms convert displacement and acceleration units to the fractional frequency shift unit. 
Under equal-arm approximation, the noise PSDs of $A_2$ and $E_2$ channels are identical: 
\begin{align}
P_{A_2, E_2}(f)  &\approx 32\sin^2 (2u) \sin^2(u) \left[\left(2 + \cos(u)\right)  P_{\rm OMS}(f) \right. \nonumber \\ 
 & \quad \ + \left. \left(6 + 4 \cos(u) + 2\cos(2u) \right) P_{\rm ACC}(f) \right], 
\end{align}
where $u \equiv 2\pi f d $, $d = 10 \ {\rm s}$ being the nominal arm-length of detector.

\end{document}